%% file: main.tex
\documentclass[a4paper,12pt, oneside]{article}
\usepackage{placeins}
\usepackage{amsmath,amssymb,amsfonts,amsthm}
\usepackage{dsfont}
\usepackage[bottom,flushmargin,hang,multiple]{footmisc}
\usepackage{setspace}
\usepackage[flushleft]{threeparttable}
\usepackage[table]{xcolor}% 
\usepackage{ragged2e} 
\usepackage{graphicx}       % Packages to allow inclusion of graphics
\graphicspath{{figures/}}
\usepackage{hyperref}                           % For creating hyperlinks in cross references
\usepackage[natbib=true,style=ext-authoryear,backend=bibtex,useprefix=true,sorting=nyt,maxnames=2, maxbibnames=99, articlein = false, date = year, giveninits=true]{biblatex}
 
\DeclareFieldFormat[article,periodical]{number}{\mkbibparens{#1}}

\DeclareFieldFormat
  [article,inbook,incollection,inproceedings,report,patent,thesis,unpublished]
  {title}{#1\isdot}

\usepackage{textcomp}                           % For single quotes
\usepackage{floatrow}                           % For image and table position
\usepackage{booktabs}                           % For tables
\usepackage[bottom]{footmisc}                   % For footnotes
\usepackage{bbm}
\usepackage{bm}

\usepackage{titling}
\usepackage{authblk}
\usepackage{etoolbox}

\usepackage[hang,flushmargin]{footmisc} 
\usepackage{mathtools}

\usepackage[table]{xcolor}  
\usepackage{tabularx,ragged2e,array}
\hfuzz=\maxdimen

\usepackage{pdfpages}
\usepackage{array}
\newcolumntype{C}[1]{>{\centering\arraybackslash}p{#1}}

\usepackage{booktabs}

\usepackage[margin=1.5in]{geometry}

\usepackage[parfill]{parskip}

\usepackage{authblk}

\usepackage{caption}
\usepackage{subcaption}
\usepackage{gensymb}
\newcommand{\dg}{\textdegree}
\title{{\Large Heat increases experienced racial segregation\\ in the United States
\vspace{0.5cm}}}
\author[1,2]{{\small Till Baldenius}}
\author[1,3,4,\thanks{\scriptsize{ \noindent Corresponding author: Nicolas Koch, Mercator Research Center on Global Commons and Climate Change (MCC), Torgauer Straße 12-15, 10829 Berlin, (\url{koch@mcc-berlin.net})}}]{{\small Nicolas Koch}}
\author[1]{{\small Hannah Klauber}}
\author[2,5]{{\small Nadja Klein\vspace{0.3cm}}}
\affil[1]{{\scriptsize Mercator Research Institute on Global Commons and Climate Change (MCC)}}
\affil[2]{{\scriptsize Chair of Statistics and Data Science (Humboldt-Universit\"at zu Berlin)}}
\affil[3]{{\scriptsize Potsdam Institute for Climate Impact Research (PIK)}} 
\affil[4]{{\scriptsize IZA Institute of Labor Economics}} 
\affil[5]{{\scriptsize Chair of Uncertainty Quantification and Statistical Learning, Research Center Trustworthy Data Science and Security (UA Ruhr) and Department of Statistics (Technische Universit\"at Dortmund)}}

\date{{\normalsize June 2023}}

\begin{document}

\maketitle

%\begin{center} 
%\vspace{0.75cm}
%WORK IN PROGRESS\\
%\end{center}
 \vspace{-1.3cm}
\input{abstract.tex}

\newpage
\input{introduction.tex}
\input{data.tex}

\input{methodology.tex}
\input{results.tex}

\input{conclusion.tex}
\newpage
\hypertarget{bibliography}{}
\renewcommand*{\bibfont}{\small}
{\scriptsize
\printbibliography
}

\newpage
%\input{appendix.tex}

% \hypertarget{references}{%
% \section*{References}\label{references}}
%\addcontentsline{toc}{section}{References}
% \renewcommand\refname{}
% \vspace{-2cm}
% \setlength{\parindent}{-0.5cm}
% \setlength{\leftskip}{0.5cm}
% \setlength{\parskip}{8pt}

% \indent
% \setlength{\parindent}{17pt}
% \setlength{\leftskip}{0pt}
% \setlength{\parskip}{0pt}
\newpage

\input{appendix.tex}

\end{document}

%% file: abstract.tex
\section*{}
%must not be longer than 900 characters
\begin{abstract} 
\noindent Segregation on the basis of ethnic groups stands as a pervasive and persistent social challenge in many cities across the globe. Public spaces provide opportunities for diverse encounters but recent research suggests individuals adjust their time spent in such places to cope with extreme temperatures. We evaluate to what extent such adaptation affects racial segregation and thus shed light on a yet unexplored channel through which global warming might affect social welfare. We use large-scale foot traffic data for millions of places in 315 US cities between 2018 and 2020 to estimate an index of experienced isolation in daily visits between whites and other ethnic groups. We find that heat increases segregation. Results from panel regressions imply that a week with temperatures above 33\dg C in a city like Los Angeles induces an upward shift of visit isolation by 0.7 percentage points, which equals about 14\% of the difference in the isolation index of Los Angeles to the more segregated city of Atlanta. The segregation-increasing effect is particularly strong for individuals living in lower-income areas and at places associated with leisure activities. Combining our estimates with climate model projections, we find that stringent mitigation policy can have significant co-benefits in terms of cushioning increases in racial segregation in the future.\\

{\footnotesize
\noindent \textbf{Keywords}: racial segregation, climate change, heat, foot traffic data, urban mobility}

\end{abstract}

%% file: introduction.tex
\hypertarget{introduction}{%
  \section{Introduction}\label{introduction}}

The negative effect of racial segregation on the socio-economic status of individuals from minority groups has been examined extensively in sociological, geographic and economic studies \cite{Massey1987, Massey1990, Cutler1997}. 
Specifically, crime and violence \citep{Massey1995, Krivo2009, Light2019}, labor market opportunities \citep{Kain1968}, education \citep{Sharkey2011}, health \citep{Collins1999, Almond2006}, and intergenerational mobility \citep{Chetty2014, marz2015intergenerational, Chetty2018b} are adversely related to racial segregation. These findings are largely limited to residency as a determinant of segregation, although daily encounters at work and leisure activities also strongly shape our experience. To better understand the dynamics of racial segregation beyond residency, segregation in the activity space, i.e.~the set of all spaces in which people move in their daily lives, has become the prime object of investigation \cite{Wong2011, Wang2018, Athey2021}. An important insight from this recent literature is that segregation is much lower when individuals spend time outside of their neighborhoods or in public spaces such as parks, restaurants and bars, or accommodation establishments \cite{Davis2019, Athey2021}. 

At the same time, a separate strand of literature on the impacts of climate change on human activity suggests that individuals use leisure and outdoor activity substitution as a behavioral adaptation strategy to extreme temperature \citep{Carman2020, BerrangFord2021}. In particular, there is evidence that outdoor leisure activity is significantly depressed when it is hot \cite{Zivin2014, Obradovich2017, Dundas2020, Fan2023}. This raises the important question whether and to what extent climate adaptation behaviors may increase the extent of segregation individuals experience. If time spent in lower-segregation outdoor leisure places, such as green or blue space, is particularly affected by heat, hot weather could increase racial segregation. However, the ultimate effect depends on how individuals substitute outdoor leisure for other activities subject to varying degrees of segregation. For instance, restaurants or bars are also associated with comparatively low levels of segregation and heat-induced substitution to these public spaces could leave overall segregation unchanged. To the best of our knowledge, the heat-segregation nexus has not been addressed as the two strands of literature on climate impacts and social segregation remained detached so far.

We contribute to resolving this deficiency and shed light on the effect of heat exposure on racial segregation by coupling foot traffic data for millions of places with spatial and temporal variations in temperature to derive the effects hot temperatures induce on segregation in the United States (US). Our large-scale foot traffic data is aggregated from app-tracking-based cell phone geolocation data collected by SafeGraph \cite{SafeGraph2022} in 315 metropolitan statistical areas (MSAs) across the US between January 2018 and March 2020. It includes weekly counts of visits and visitors to almost eight million points of interest (POI) by the census block groups (CBGs) of visitors' residences, for which we infer the racial composition using census data. Building on prior contributions \cite{Gentzkow2011, Athey2021}, we compute an index of weekly experienced isolation in visits. The index provides an estimate of how much less likely it is for a {non-\textit{White}} visit to coincide with a {\textit{White}} visit than it is for a {\textit{White}} visit. %\footnote{\textit{White} and non-\textit{White} visits denote visits that originate from CBGs, where the majority of residents identify as White or as other than White, respectively.}. 
We infer the heat exposure of visits within MSAs using daily data from the Parameter-elevation Regressions on Independent Slopes Model (PRISM) climate mapping system \cite{PRISMClimateGroup2022}. 

Our empirical strategy follows the well-established climate–econometric literature \citep{Dell2014, Carleton2016, Hsiang2016} and applies fixed-effects panel regression models to exploit week-to-week variation of temperature and visit isolation within MSAs (see \hyperref[sec:femod]{Model Specification}). It allows us to flexibly control for time-constant differences between MSAs, weekly nationwide shocks, and state-specific trends and seasonality at the monthly level in visit isolation and temperature. We argue that temperature variation net of these fixed effects is plausibly random, allowing causal interpretation of statistically significant estimates. Compared with survey data that is often limited by self-reporting bias and a small set of respondents, our app-tracking-based data provided by SafeGraph give a more objective and representative picture of human activity under heat exposure. We show that the data are representative in terms of sex, age, education, and race, while wealthy households are slightly over-represented (see Appendix Figs.~\ref{fig:app1}-\ref{fig:app5}).

%% file: data.tex
\hypertarget{data}{%
\section{Data}\label{data}}
\vspace{0.2cm}

\subsection*{Foot-traffic Data}

We use data on about eight million POI, that is, geographic locations such as restaurants, grocery stores, urban transit systems, and nature parks, provided by the US company {SafeGraph}. The 2017 North American Industry Classification System (NAICS) codes associated with the respective establishments allow for a detailed categorization of POI.

Information on foot traffic to the POI is provided with the \textit{Weekly Patterns} dataset, which is available as of January 1, 2018.  When accessed in March 2022, \textit{Weekly Patterns} covered an average of 0.052 devices per person in the MSAs of the CONUS. {SafeGraph} combines geometric information about places with location data from smartphone application users to determine visits to POI. When using an application whose provider partners with {SafeGraph}, geolocation pings that indicate the location of a device when used for the purposes of the application are collected via app-tracking. If such a ping is within a building's geographic footprint and the next ping at least four minutes apart is in the same polygon, {SafeGraph} calls this a visit. These aggregations include the numbers of visits and visitors. For each POI, the number of visitors is broken down by the visitors' home CBGs.

To determine the origin neighborhood of visitors, {SafeGraph} infers a device's home location from its primary nighttime geohash7 (a geographical unit from a grid of approximately 500 $\times$ 500 feet) determined from at least six weeks of tracking. 
This geohash7 is then mapped to the corresponding CBG, the highest geographic resolution for which demographic data are available from the census. 
Linking SafeGraph data with demographic information from the census, we conclude that the data is representative of sex, age, education, and race, while wealthy households are slightly over-indexed (see Appendix Figs.~\ref{fig:app1}-\ref{fig:app5}).

{SafeGraph} data are widely used in academic research, particularly in articles related to the COVID-19 pandemic \citep{Allcott2020, Benzell2020, Chang2021, Ma2021, Li2022, Jay2022}. Specifically, these authors examine how social segregation was affected by the pandemic \cite{Park2021, Li2022}, whether racial segregation differs between students and adults \cite{Cook2022}, and they explore differences in movement patterns of population sub-groups \cite{Chen2020, Prestby2020}.

\subsection*{Weather Data and Climate Projections} \label{sec:wd}

The Parameter-elevation Regressions on Independent Slopes Model (PRISM) climate mapping system \citep{PRISMClimateGroup2022}, developed in the 1990s, provides official climate data of the CONUS for the Department of Agriculture of the United States. One of the PRISM products is the daily time series of CONUS weather variables, including minimum and maximum temperature as well as precipitation. The estimates are available from 1981 and are gridded at a resolution of approximately 4km. The focus of this product is to obtain the best possible estimates regardless of temporal consistency and thus uses all weather stations available at the time of estimation \citep{PRISMClimateGroup2022}. Using the R handbook that is part of the supplementary material to ref. \cite{Ortiz2021}, we aggregate the gridded data to the MSA level. We weight by grid-cell population \citep{Burke2018}, for which we use gridded block-level population data from ref. \cite{Falcone2016}. This method of spatial aggregation reflects the weather condition to which the average person within an MSA was exposed. We measure precipitation in millimeters and maximum temperature in degrees celsius. We consider maximum temperature because it is usually the primary measure of how hot a day is, such as in weather forecasts. 

For future climate projections, we consult bias-corrected daily maximum temperatures from the Geophysical Fluid Dynamics Laboratory's Princeton Earth System Model (GFDL-ESM4) \citep{lange_buechner_2021}. The model belongs to the global climate models in the sixth phase of the Coupled Model Intercomparison Project (CMIP-6). We obtain the data for the year 2050 for a best-case scenario with strict climate policy (SSP1/RCP2.6) and a worst-case scenario without emissions abatement (SSP5/RCP8.5). To aggregate the projection data to the MSA level, we apply the same procedure as to the PRISM data.

%% file: methodology.tex
\hypertarget{methodology}{%
\vspace{-0.3cm}
\section{Methodology}\label{methodology}}

\subsection*{Visit Isolation Index Estimation}
\label{sec:viindexestimation}

For derivation of an index of racial segregation, we divide visitors into the groups \textit{White} and non-\textit{White} based on their home CBG \cite{Athey2021}. Using 2010 census data, we classify a CBG as \textit{White} if the majority of the population of that CBG identifies as White, and otherwise as non-\textit{White}. Based on this distinction, visits originating from a particular CBG are referred to as \textit{White} or non-\textit{White}. Alternatively, race could be assigned to individual visits by stochastic imputation based on the racial composition of the origin CBG. However, this could lead to an underestimation of the extent of segregation because it implies that residents of the same CBG are equally likely to visit POI regardless of self-reported race \citep{Athey2021}.

While our goal is to best reflect individual experiences in the segregation measure used for our analysis, devices are not identified across locations in {Weekly Patterns}. This makes it impossible to determine the actual activity space of any of the individuals. Therefore, our focus is on segregation in visits rather than experienced segregation. Experienced segregation is driven by two factors. One is the set of places individuals visit. If members of two groups systematically choose different places, this contributes to segregation. This factor may reflect a preference for places close to home or work if those places are themselves segregated, it may reflect particular preferences that differ systematically by group, such as a preference for going to the movies over going to the theater, and it may reflect a possibly unconscious preference for places where one's group is in the majority. The second factor is the number of places an individual chooses to visit in a given time. The latter factor cannot be represented by our data. Therefore, we estimate how segregated visits are rather than focusing on individuals in our analysis.
Regarding the choice of space within which we want to estimate the extent of segregation, we decide to focus on MSAs, as is common in previous contributions \cite{Cutler1997, Echenique2007, Athey2021}. 

We aim to construct an estimator based on the isolation index that measures segregation of visits to POI within MSAs. To this end, we define $a_l^V$, $b_l^V$ and $t_l^V$ as the number of group $a$, group $b$ and total visits to POI $l \in L_m$, respectively, where $L_m$ is the set of POI in area $m$. For some of these visits, people travel from outside the area in which the POI is located. Therefore, we distinguish between visits from within an MSA and visits from outside it. Our goal is to measure segregation from the perspective of those who live within the area, but to consider that this perspective is also influenced by those who visit POIs from outside the area. With that in mind, we define \textit{exposure} as
 
\begin{equation}
	E_{a \times b, m} = \sum_{j\in A_m^V}  \frac{1}{|A_m^V|}\frac{b_{l(j)}^V}{t_{l(j)}^V},
	\label{eq:pindad}
\end{equation}
where $A_m^V$ denotes the set of visits of individuals from group $a$ that have home locations in area $m$. Defining $b_{l,m}^V$ as the visits count to location $l$ of individuals from group $b$ that live in area $m$, we can reformulate Eq.~\ref{eq:pindad} as 
\begin{equation}
	E_{a \times b, m} = \sum_{l \in L_m}  \frac{a_{l,m}^V}{|A_m^V|}\frac{b_l^V}{t_l^V}.
	\label{eq:pad}
\end{equation}
Note that the first factor only considers visitors that reside in area $m$, whereas the second factor accounts for all visitors with home locations in the United States. Substituting the \textit{exposure} measure $E$ into the isolation index \cite{Athey2021} yields the visit isolation index
\begin{align}
	VI_m &= \sum_{j\in B_m^V}  \frac{1}{|B_m^V|}\frac{b_{l(j)}^V}{t_{l(j)}^V} - \sum_{j\in A_m^V}  \frac{1}{|A_m^V|}\frac{b_{l(j)}^V}{t_{l(j)}^V} 	\label{eq:viso} \\
	&= \sum_{l \in L_m}  \frac{b_{l,m}^V}{|B_m^V|}\frac{b_l^V}{t_l^V} - \sum_{l \in L_m}  \frac{a_{l,m}^V}{|A_m^V|}\frac{b_l^V}{t_l^V},	\label{eq:visol}
\end{align}
which is our preferred estimator of the extent of segregation. For this purpose, we apply the distinction of non-\textit{{White}} and {\textit{White}} visits defined earlier. We call these groups $nw$ and $w$, respectively, and the visit isolation index becomes
\begin{equation}
	VI_m = \sum_{l \in _m}  \frac{w_{l,m}^V}{|W_m^V|}\frac{w_l^V}{t_l^V} - \sum_{l \in L_m}  \frac{nw_{l,m}^V}{|NW_m^V|}\frac{w_l^V}{t_l^V}. \label{eq:visolw}
\end{equation}
It follows that, on scale of zero to unity, the visit isolation index $VI_m$ estimates the extent to which the population that identifies as {{White}} is exposed to itself more than the population that identifies as non-{{White}} is exposed to it.

Unfortunately, we only know how many different individuals from a given home CBG stayed at a given POI at least once in a given week, but not whether they visited the POI multiple times and, if so, how many times. It follows that we can only estimate the visit isolation index of Eq.~\ref{eq:visolw}. To estimate visits by group based on the available information, we multiply the number of visitors at a given location, $nw_l^N$ and $w_l^N$, by the visits per visitor at that location, $t_l^V/t_l^N$. We estimate the total number of visits by the two groups in area $m$ in a similar manner as 
\begin{equation}
	\widehat{|NW_m^V|} := \sum_{l \in L_m} \widehat{nw}_{l,m}^V := \sum_{l \in L_m} {nw_{l,m}^N}\frac{t_l^V}{t_l^N},
\end{equation}
where is $\widehat{|NW_m^V|}$ defined accordingly.
Finally, our estimator of visit isolation is
\begin{equation}
	{VI}^\text{est}_m = \sum_{l \in L_m}  \frac{\widehat{w}_{l,m}^V}{\widehat{|W_m^V|}}\frac{\widehat{w}_l^V}{t_l^V} - \sum_{l \in L_m}  \frac{\widehat{nw}_{l,m}^V}{\widehat{|NW_m^V|}}\frac{\widehat{w}_l^V}{t_l^V}
    \label{eq:visolwest}
\end{equation}

Fig.~\ref{fig:tm_vi} shows averages of weekly visit isolation estimates from January 1, 2018--March 8, 2020. We choose this period to precede potentially COVID-19-induced changes in mobility behavior to eliminate a potential source of confounding. There are 65 MSAs with insufficient non-{\textit{White}} visits that are colored gray on the map, because they have an insufficient number of recorded visits from non-\textit{White} CBGs, which are excluded from the analysis, limiting our sample to 315 MSAs.

\subsection*{Model Specification}\label{sec:femod}

To examine the relationship between heat and the visit isolation index by exploiting temporal variation, we fit the fixed effects model
\begin{equation}
	VI^\text{est}_{i\tau} = H_{i\tau} \beta + P_{i\tau}\rho  + \mu_i + \delta_\tau + \eta_{sm} +  \epsilon_{i\tau}, \label{eq:estimate}
\end{equation}
where $i = 1, \dots, N=315$ and $\tau = 1, \dots, T = 114$ denote MSA and week, respectively. $VI^\text{est}_{i\tau}$ indicates the estimator of the visit isolation index as denoted by Eq.~\ref{eq:visolwest} and $H_{i\tau}$ and $P_{i\tau}$ are $(1 \times k_H)$ and $(1 \times k_P)$ vectors of daily maximum temperature bins and daily precipitation bins, respectively. The associated coefficients are denoted by the  vector $\beta\in\mathds{R}^{k_H \times 1}$ and $\rho\in\mathds{R}^{k_P \times 1}$, with $\beta$ being the main coefficient of interest. There are MSA fixed effects $\mu_i$, week fixed effects $\delta_\tau$ and a third set of fixed effects $\eta_{sm}$ that account for time trends specific to the states $s$ and months $m$ in our preferred specification. Finally, $\epsilon_{i\tau}$ denotes the error term.

The term fixed effect refers to the fact that we model $\mu_i$ and $\xi_{i\tau} :=[\delta_\tau, \eta_{i\tau}]$ as parameters and thus allow for arbitrary correlation with ${X}_{i\tau} := [H_{i\tau}, P_{i\tau}]$. 

In this case, assuming
\begin{equation}
	E[\epsilon_{i\tau}|X_{i1}, \dots, X_{iT},  \mu_i, \xi_{i1}, \dots, \xi_{iT}] = 0, \hspace{1em}\tau = 1, \dots, T, \label{eq:re_a1}
\end{equation}
is sufficient to obtain strict exogeneity of the form
\begin{equation}
E[\epsilon_{i\tau}|X_{i1}, \dots, X_{iT}] = 0, \hspace{1em}\tau = 1, \dots, T. \label{eq:fe_se}
\end{equation}
Since they were prominently applied in a climate-agriculture \cite{Deschenes2007}, panel data approaches employing fixed effects have been widely used in the climate econometrics literature. In the specific case of regressions involving weather variables, deviations of weather variables from their unit-specific averages are presumed to be as good as random after controlling for temporal shocks \citep{Deschenes2007}. This plausibly random weather variation net of fixed effects establishes quasi-experimental conditions, thus allowing for the identification of causal effects \citep{Dell2014}. 

When employing fixed effects, time-constant characteristics of the units that may correlate with both the outcome and the variables of interest, and may thus bias the estimates, do not need to be explicitly modeled \citep{Wooldridge2010, Hsiang2016}. Nevertheless, there could be time-varying factors that correlate with $X_{i\tau}$ after controlling for temporal shocks via $\delta_\tau$ and $\eta_{i\tau}$. The distribution of $X_{i\tau}$ is determined by geophysical processes though and therefore other relevant climatic variables not characterized by $X_{i\tau}$ already are the main concern here \citep{Hsiang2016}. Including additional non-climatic variables could itself be a problem, as these could themselves be affected by weather variation and therefore bias estimates by over-controlling \citep{Dell2014}. We argue that the deviation of weather variables from their MSA-specific averages after controlling for country-wide weekly and state-wide monthly shocks is as good as random and thus uncorrelated with other determinants of segregation. If this is the case and the model is correctly specified, then strict conditional exogeneity as in Eq.~\ref{eq:fe_se} holds in our preferred specification, making the estimator $\hat{\beta}$ unbiased and causally interpretable.

%\subsection*{Climate Projections}

%To project how the number of encounters of non-white individuals with white individuals would change under different climate scenarios, we proceed as follows: First, we calculate the weekly difference in the number of days per temperature bin between the two climate scenarios in 2050 and the reference period from 1987 to 2017. Second, we multiply the temperature differences with the estimated coefficients $\hat{\beta}$ from Eq.~\ref{eq:estimate} and the 
%total volume of visits. To obtain a population-wide change in the number of encounters per MSA, we multiply the calculated quantities with the ratio of MSA population size over the number of devices recorded by Safegraph (see Appendix Table~\ref{tab:clim_proj}). To derive the individual-level change in the number of encounters, we alternatively divide the quantities by the number of registered devices (see Fig.~\ref{fig:ssp}).

%% file: results.tex
\hypertarget{results}{%
\section{Results}\label{sec:res}}
\vspace{0.2cm}

\begin{figure}[t!]
     \centering
		\includegraphics[width = \linewidth]{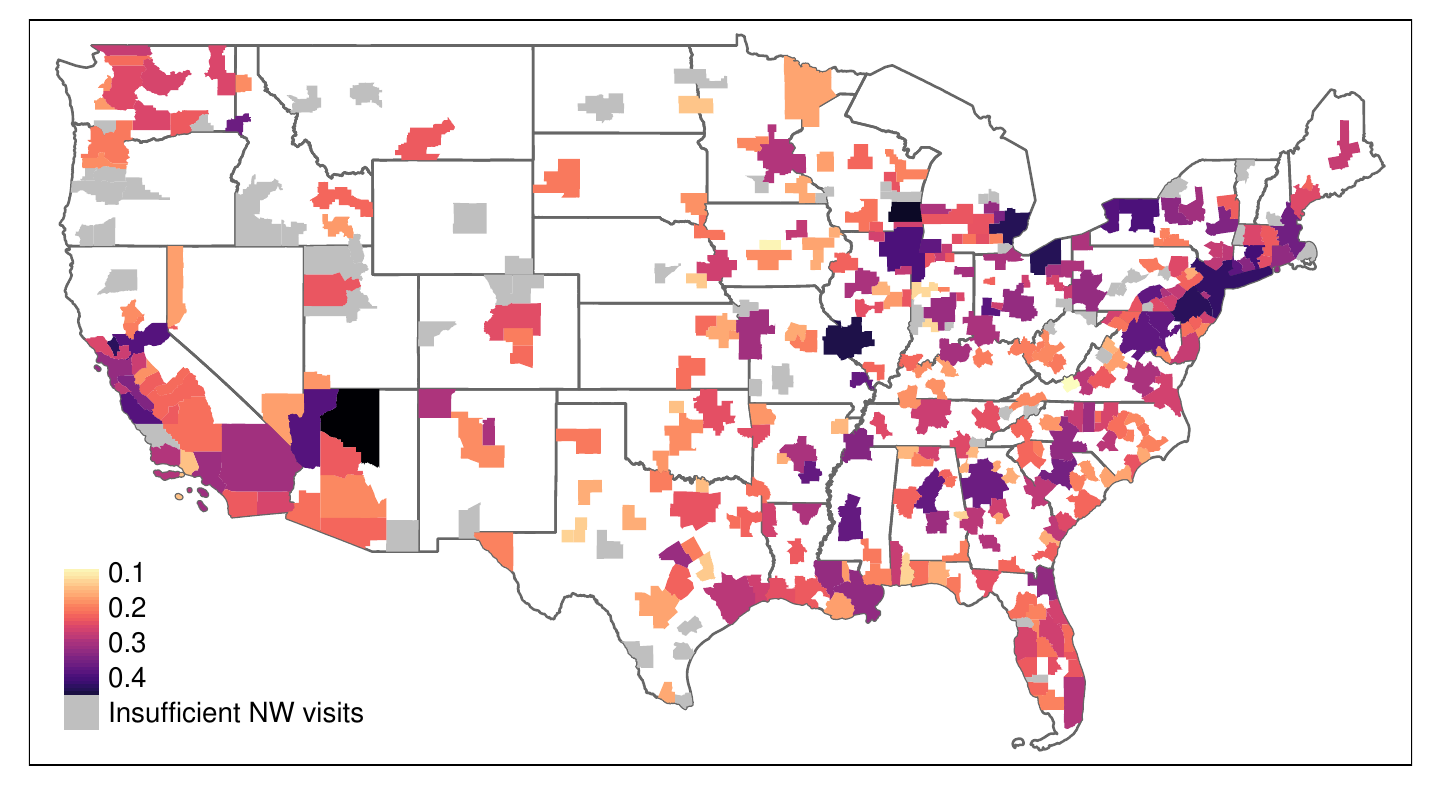}
		\caption{Average visit isolation index} \justifying{\footnotesize \noindent Average visit isolation index estimates for the MSAs in the CONUS, January 1, 2018--March 8, 2020. Light-colored areas indicate lower estimates of isolation. Gray-shaded areas did not have any visits labeled {non-{\textit{White}}} in at least one time period and thus isolation could not be estimated in all time periods.}
		\label{fig:tm_vi}
 \end{figure}

Our outcome of interest is a visit isolation index that measures racial segregation in visits to POI within MSAs (see \hyperref[sec:viindexestimation]{VI Index Estimation}). Fig.~\ref{fig:tm_vi} and Fig.~\ref{fig:vi_var} show the average visit isolation index across 315 MSAs and how visit isolation varies over time in the three most populous MSAs, respectively. Isolation tends to be higher in the more populous MSAs compared with the smaller MSAs. The visit isolation index of about 0.4 in New York implies that a visit from a majority White CBG is 40 percentage points more likely to coincide with another visit from a majority White CBG than a visit from a majority non-White CBG. 

Fig.~\ref{fig:main} shows the estimated effect of maximum temperatures on visit isolation. We form 5\dg C temperature bins and count the number of days within a week on which the maximum temperature was in a given bin to model heat exposure. Thus, we model the temperature-segregation relationship flexibly. Each of the individual coefficients is interpretable as the change in estimated visit isolation associated with an additional day of maximum temperature in the respective bin relative to the reference bin of 20\dg C to 25\dg C.

\subsection*{Main Results}

\begin{figure}[t!]
	\centering
		\includegraphics[width = \linewidth]{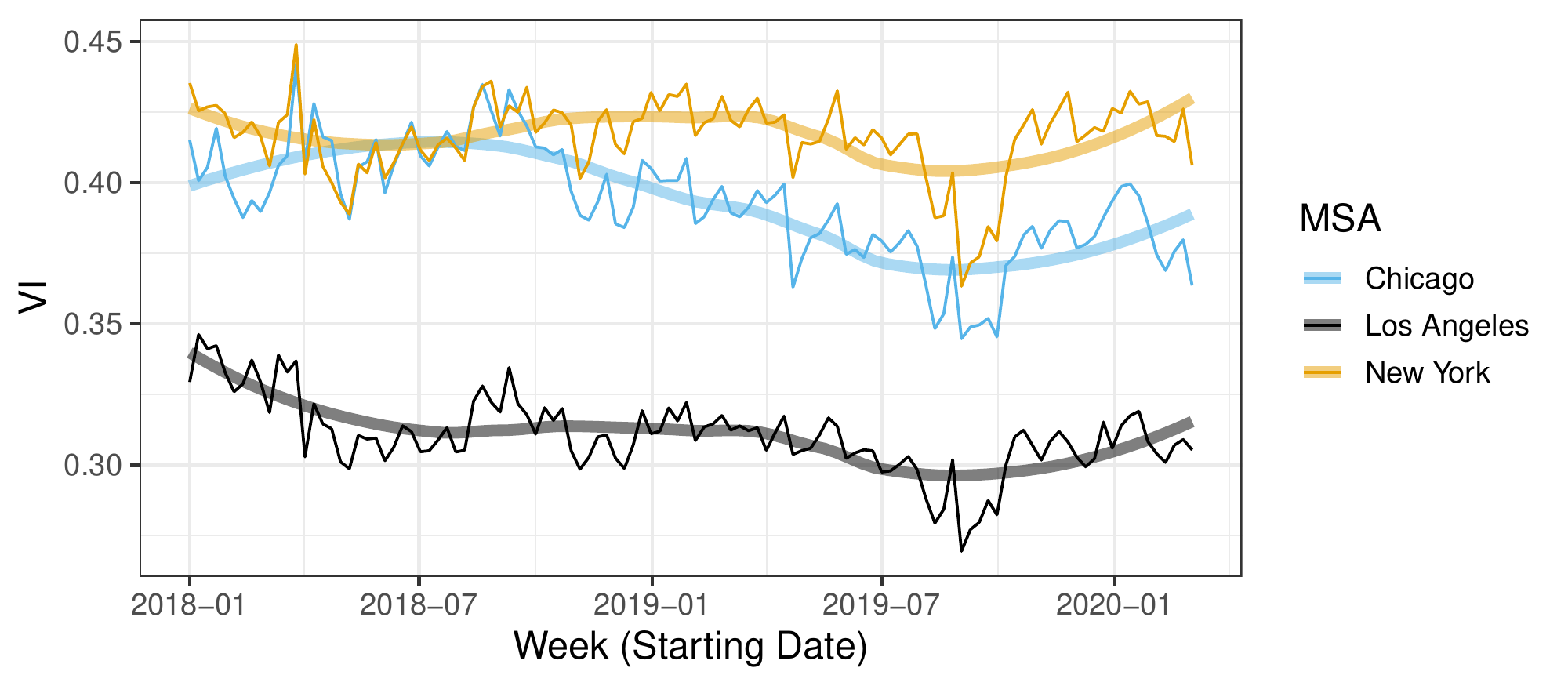}
		\caption{Visit isolation index (VI) for the three most populous MSAs}
  \justifying{\footnotesize \noindent
  Visit isolation index (VI) estimates from January 1, 2018 -- March 8, 2020 for the three most populous MSAs in the United States. The blue, black, and yellow lines refer to the MSAs of Chicago, Los Angeles, and New York, respectively. The smooth lines depict a local quadratic polynomial regression fit with span parameter $\alpha = 0.6$. }
		\label{fig:vi_var}
  \end{figure}

\begin{figure}[t!]
 \begin{subfigure}[b]{\linewidth}
	\centering
		\includegraphics[width = \linewidth]{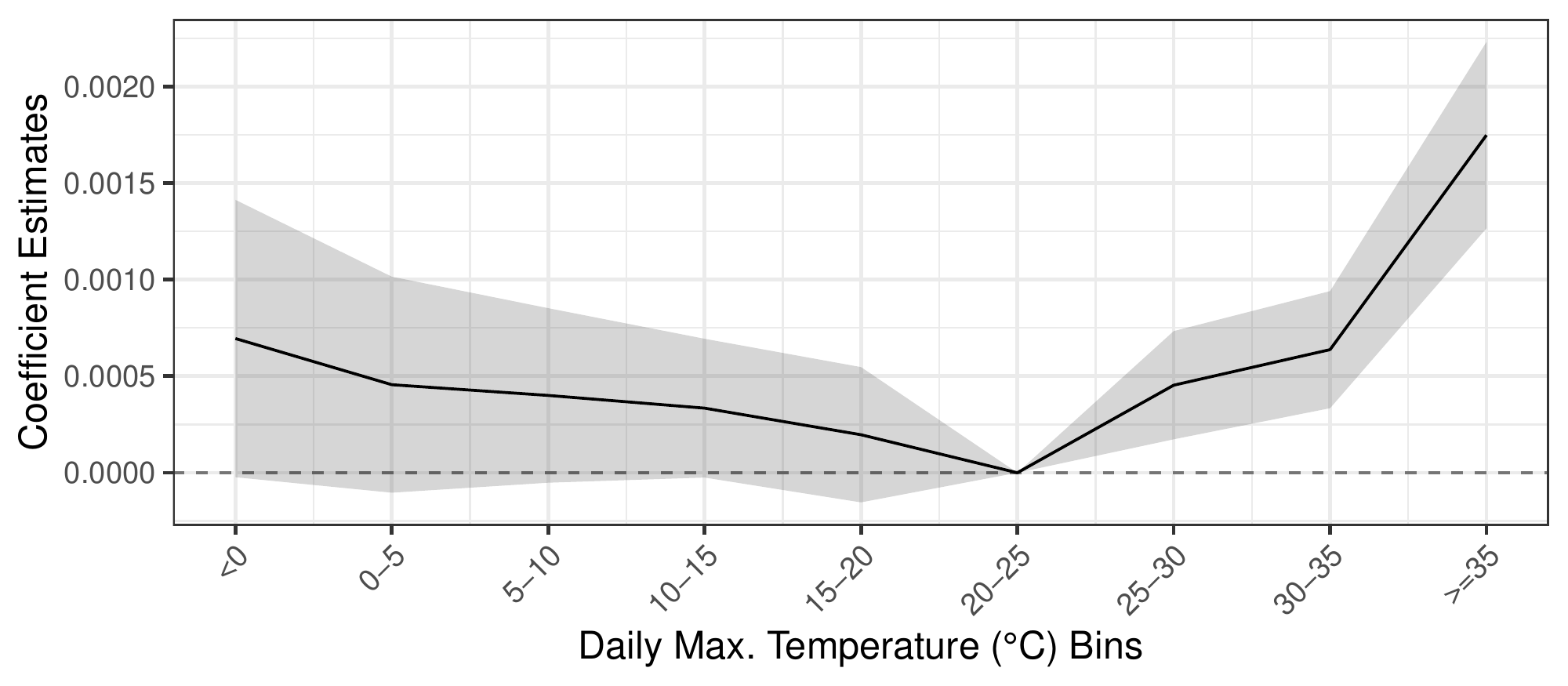}
		\caption{\footnotesize{Coefficients estimated by regressing visit isolation on daily maximum temperature bins using a fixed effects model. The 20\textdegree C to 25\textdegree C bin serves as a reference bin and precipitation is included as a control variable. MSA, week, and state-by-month fixed-effects are employed and observations are weighted by MSA population size. Shaded areas depict asymptotic 95\% confidence intervals based on Conley-HAC standard errors.}}
	\label{fig:main}
   \end{subfigure}\\
   
    \hfill
\begin{subfigure}[b]{\linewidth}
  \includegraphics[width = \linewidth]{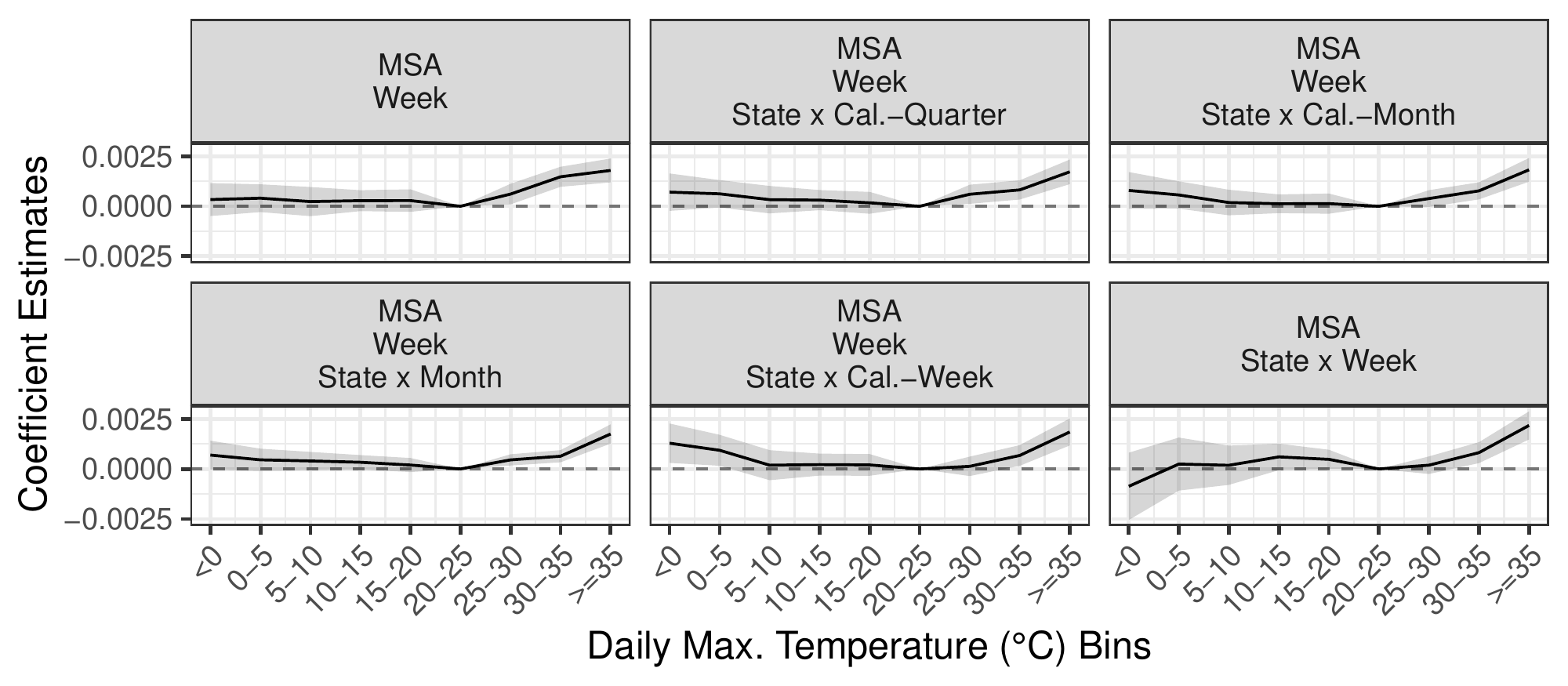}
  \caption{ \footnotesize{Re-estimations of the relationship between visit isolation and daily maximum temperature with alternative fixed effects specifications as indicated by the subheadings of each panel. The preferred specification from Fig.~\ref{fig:main} (bottom left) is compared to specifications neglecting subregional time trends (top left panel), controlling for statewide seasonality in calendar quarters (top middle), calendar months (top right), and calendar weeks (bottom middle), and statewide weekly time trends (bottom right).
  }}\label{fig:fe}
    \end{subfigure}	
   \caption{The relationship between visit isolation and daily maximum
temperature}
\end{figure}

We find a U-shaped relationship between temperature and visit isolation. While the effects of additional cold days on our visit isolation index are positive but statistically insignificant, our estimates suggest a strong effect of days with high temperature on visit isolation. Fig.~\ref{fig:main} reveals that an additional very hot day (up to 35\dg C and higher) is associated with a statistically significant 0.17 percentage point increase in the isolation index relative to an additional mild day (up to 20\dg C to 25\dg C). In other words, we estimate that in order for the probability of a {non-\textit{White}} visit to coincide with a {\textit{White}} visit to cumulatively decrease by more than one percentage point relative to the probability of a {\textit{White}} visit to coincide with another {\textit{White}} visit, six additional very hot days (16 days of up to 30\dg C to 35\dg C; 23 days of up to 25\dg C to 30\dg C) are required. In order to help building some intuition about the magnitude of this effect, we provide a few benchmarks. In August 2018, Los Angeles experienced a week of average maximum temperatures above 33\dg C. Our estimates imply that the heat effect of such a week entails an upward shift of visit isolation by 0.7 percentage points. This shift equals about 14\% of the difference in visit isolation of Los Angeles to Atlanta, 32\% of the difference to Miami, and 80\% to San Francisco. As an alternative benchmark, we compare our results to segregation trends. In Los Angeles, our visit isolation index shows a decreasing trend over the sample period (Fig.~\ref{fig:vi_var}). The positive effect we find for the heat week in the city is approximately thirty times the decrease in the trend.

\subsection*{Robustness Checks}

To control for spatial and temporal confounders of visit isolation, the inclusion of MSA and week-specific fixed effects represents a minimum requirement for our model specification. These fixed effects restrict the model to exploit only variation within MSAs and allow for flexibly absorbing nationwide time trends in the isolation index as well as effects linked to specific occasions such as national holidays, elections or big sports events. To account for potential regional time-varying sources of confounding that could be correlated with both weather and the isolation index, we include additional fixed effects at the state-month level. While we believe this fixed effects choice to be most suitable to recover the parameter of interest, the degree of stringency may be chosen differently. For this reason, Fig.~\ref{fig:fe} shows our preferred fixed effects specification in the bottom left panel along with five alternative fixed effects specifications, all of which include at minimum an MSA and a week-specific fixed effect. The panels from left to right and top to bottom are ordered by increasing level of stringency in the fixed effects. In all specifications, coefficient estimates and standard errors for high temperature bins are very robust compared to the preferred specification. Most specifications also indicate a positive but weaker effect of cold temperatures; however, it is mostly statistically insignificant. Results are also robust to computing clustered standard errors instead of Conley-HAC standard errors (see Appendix Fig.~\ref{fig:se}) and to changing the geographic area of reference from MSAs to urban areas as delineated for the US census (see Appendix Fig.~\ref{fig:UA}).

\subsection*{Heterogeneity Analysis}

\begin{figure}[b!]
		\begin{subfigure}[b]{0.325\linewidth}
			\includegraphics[width=\linewidth]{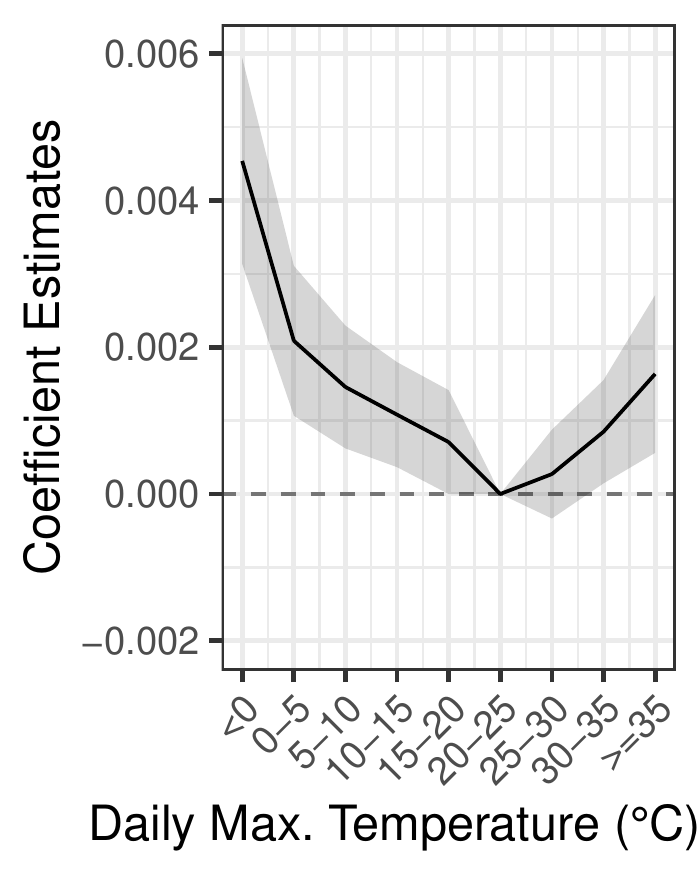}
			\caption{Outdoor Leisure}
			\label{fig:outdoor}
		\end{subfigure}
		\hfill
		\begin{subfigure}[b]{0.325\linewidth}
			\includegraphics[width=\linewidth]{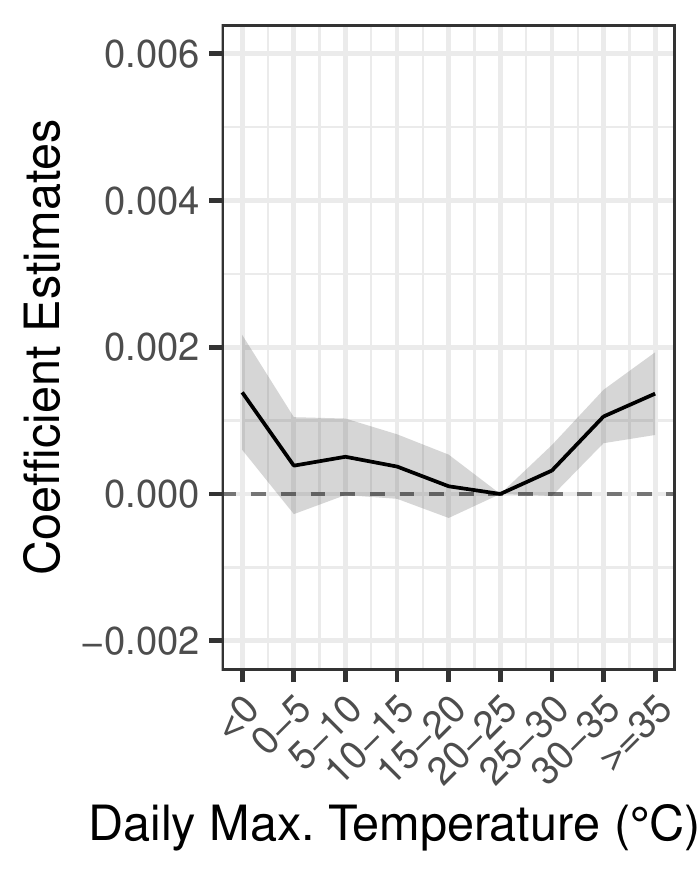}
			\caption{Indoor Leisure}
			\label{fig:indoor}
		\end{subfigure}
		\hfill
		\begin{subfigure}[b]{0.325\linewidth}
			\includegraphics[width=\linewidth]{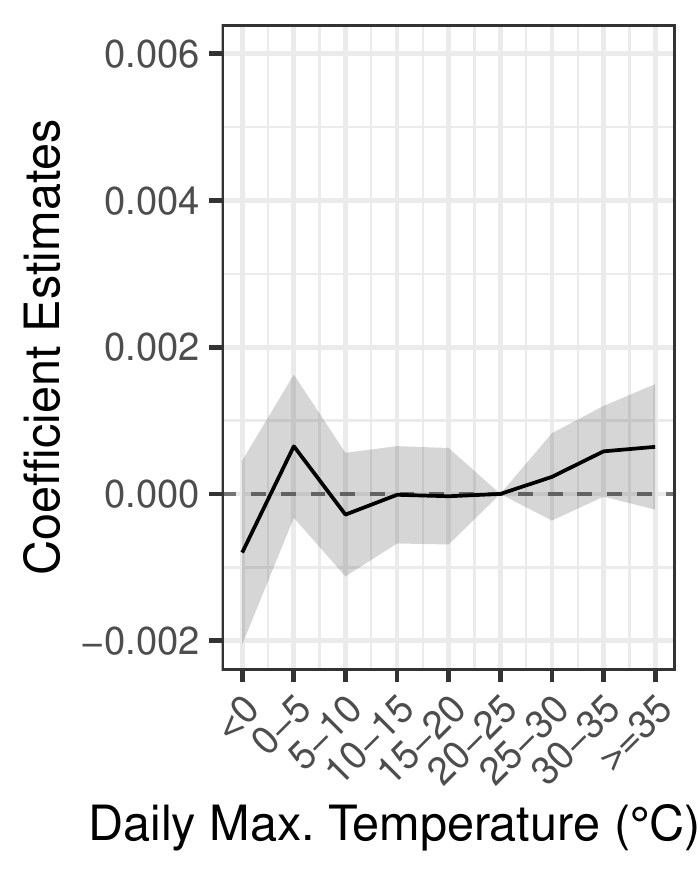}
			\caption{Grocery Stores}
			\label{fig:grocery}
		\label{fig:naics}
  \end{subfigure}\\
    \caption{The relationship between visit isolation and daily maximum temperature across different places of daily activity}
\justifying \noindent \footnotesize{The three panels present the relationship between visit isolation and daily maximum temperature estimated at outdoor POI associated with leisure activities (\ref{fig:outdoor}), indoor POI associated with leisure activities (\ref{fig:indoor}), and grocery stores (\ref{fig:grocery}). All estimates are based on the regression specification presented in Fig.~\ref{fig:main}.}
  \end{figure}

\begin{figure}[b!]
 \begin{subfigure}[b]{\linewidth}
     \centering
		\includegraphics[width = \linewidth]{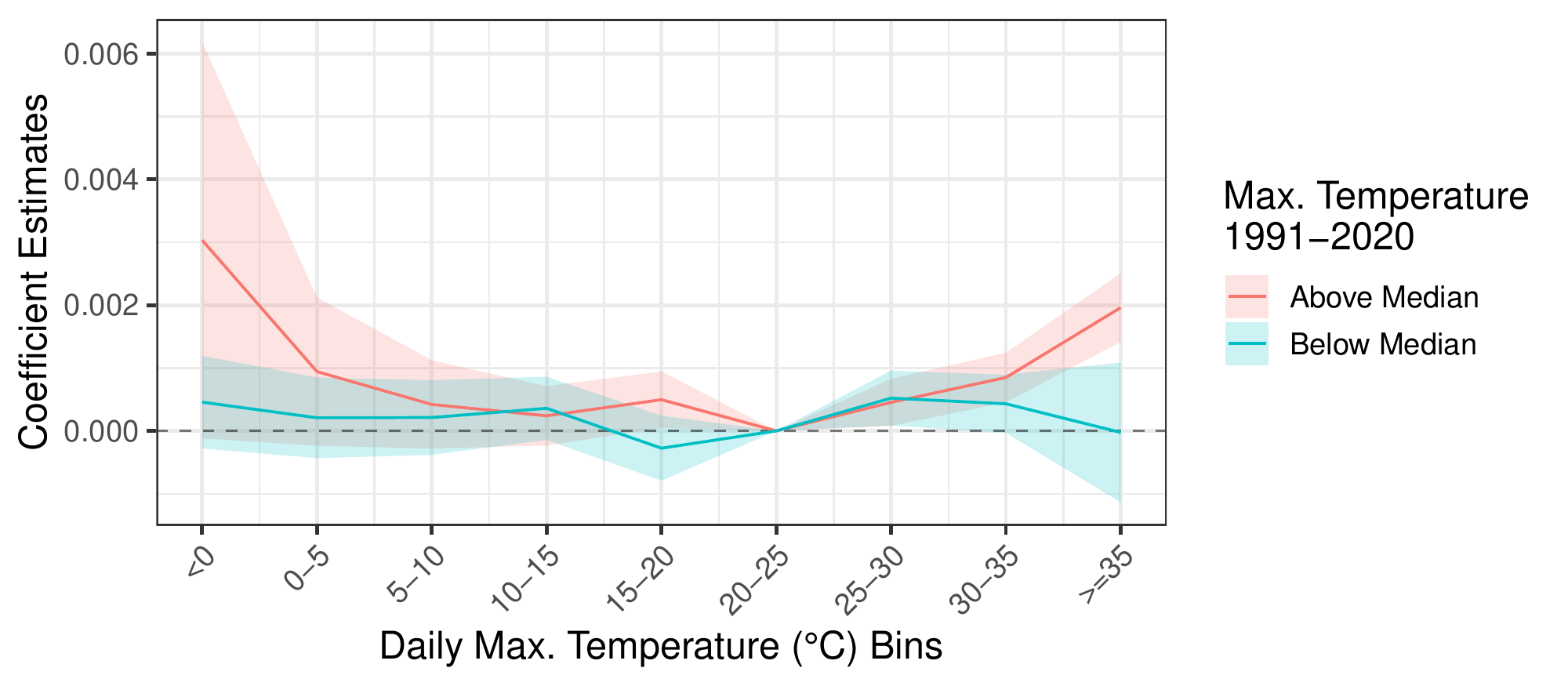}
		\caption{\footnotesize{ MSAs below and above median 30-year normal maximum temperature}}
		\label{fig:tempsplit}
 \end{subfigure}
\begin{subfigure}[b]{\linewidth}
	\centering
		\includegraphics[width = \linewidth]{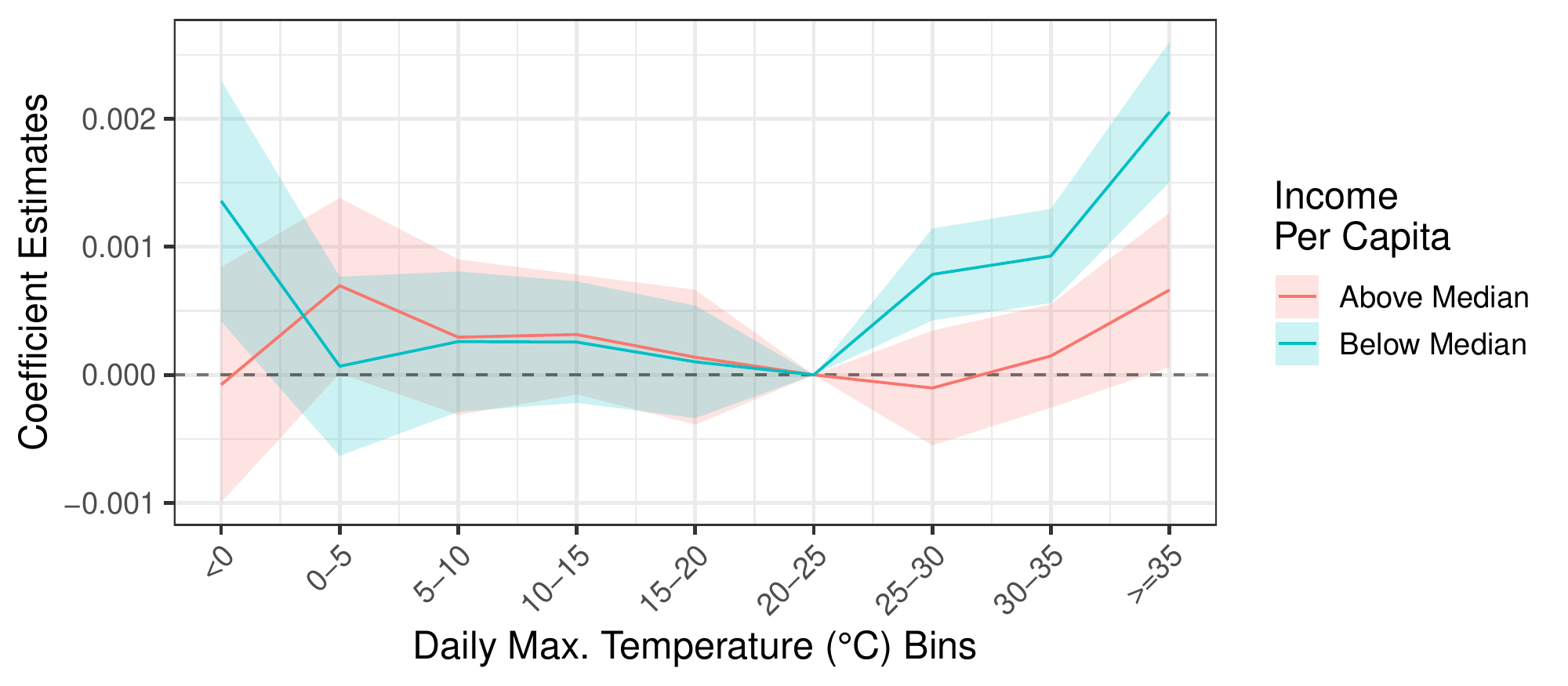}
		\caption{\footnotesize{MSA below and above median household income per capita}}
		\label{fig:incsplit}
  \end{subfigure}
  \caption{Heterogeneity in the relationship between visit isolation and daily maximum temperature}
  \justifying \noindent \footnotesize{The panels present the relationship between visit isolation and daily maximum temperature when interacting weather variables with an indicator of being above the population-weighted median of 30-year normal maximum temperature (\ref{fig:tempsplit}) and household income per capita (\ref{fig:incsplit}). All estimates are based on the regression specification presented in Fig.~\ref{fig:main}.}
\vspace{-0.5cm}
\end{figure}

To gain an initial understanding of potential mechanisms driving the heat-isolation relationship, we analyze how the effect of heat on visit isolation is heterogeneous across three categories of places: outdoor POI associated with leisure activities, indoor POI associated with leisure activities, and grocery stores (Figs.~\ref{fig:outdoor}--\ref{fig:grocery}). At outdoor places, both very low and very high temperatures are associated with higher isolation compared to moderate temperatures, with the effect of cold weather being particularly strong. Indoor places show a similar relationship between high temperatures and visit isolation, while low temperatures do not have nearly as significant an effect. While decreases in leisure and outdoor activity related to hot weather have been shown in previous studies \cite{Zivin2014, Obradovich2017, Dundas2020, Fan2023}, we provide new evidence documenting that this adaptation to heat induces a higher degree of visit isolation. In contrast, grocery stores exhibit fairly constant levels of isolation across the weather distribution, potentially due to a lower elasticity in demand for groceries.

We also examine whether the relationship between temperature and isolation varies with MSA population size. In general, segregation patterns can vary substantially across cities. While larger cities tend to have a more diverse population overall, they may be more segregated across neighborhoods. These baseline differences may also affect how strong the potential for temperature-induced changes in segregation patterns is. For instance, moving between areas with different racial and ethnic compositions during periods of heat could constitute a greater effort in larger cities. We find that the effect of heat is strongest in MSAs of the second population size quartile, corresponding to between 300 thousand and 700 thousand inhabitants (see Appendix Fig.~\ref{fig:popsplit}). However, uncertainty in the estimates prevents us from identifying clear patterns of effect heterogeneity. 

As a final heterogeneity analysis, we interact maximum temperature with an indicator of being above the population-weighted median of MSA-level 30-year average maximum temperature in a varying-coefficient model. 
Fig.~\ref{fig:tempsplit} shows that there is a statistically differential effect only for the association between isolation and heat of at least 35\dg C. On such extremely hot days only MSAs in hotter climates exhibit increased isolation. This suggests that there is limited to no adaptation to hot temperatures in terms of maintaining diversity of potential encounters in cities where heat is more common. 
Fig.~\ref{fig:incsplit} shows the coefficient estimates when interacting the maximum temperature bins with an indicator for a per capita income that is above or below the population-weighted median. The estimates reveal a strong heat effect on visit isolation in poorer MSAs, whereas it is statistically insignificant in richer ones.

\subsection*{Heat-induced segregation in a changing climate}

\begin{figure}[b!]
    \begin{subfigure}[b]{\linewidth}
        \centering
		\includegraphics[width = \linewidth]{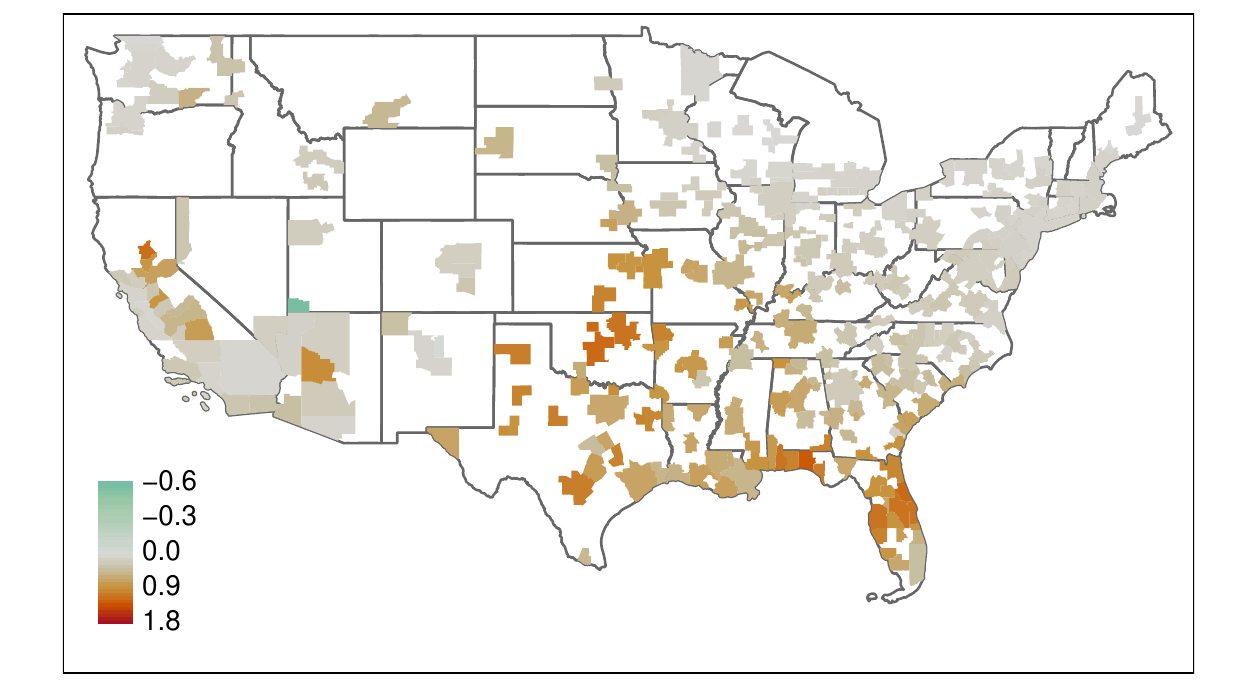}
		\caption{Under SSP3}
		\label{fig:ssp3}
    \end{subfigure}
    \begin{subfigure}[b]{\linewidth}
        \centering
        \includegraphics[width = \linewidth]{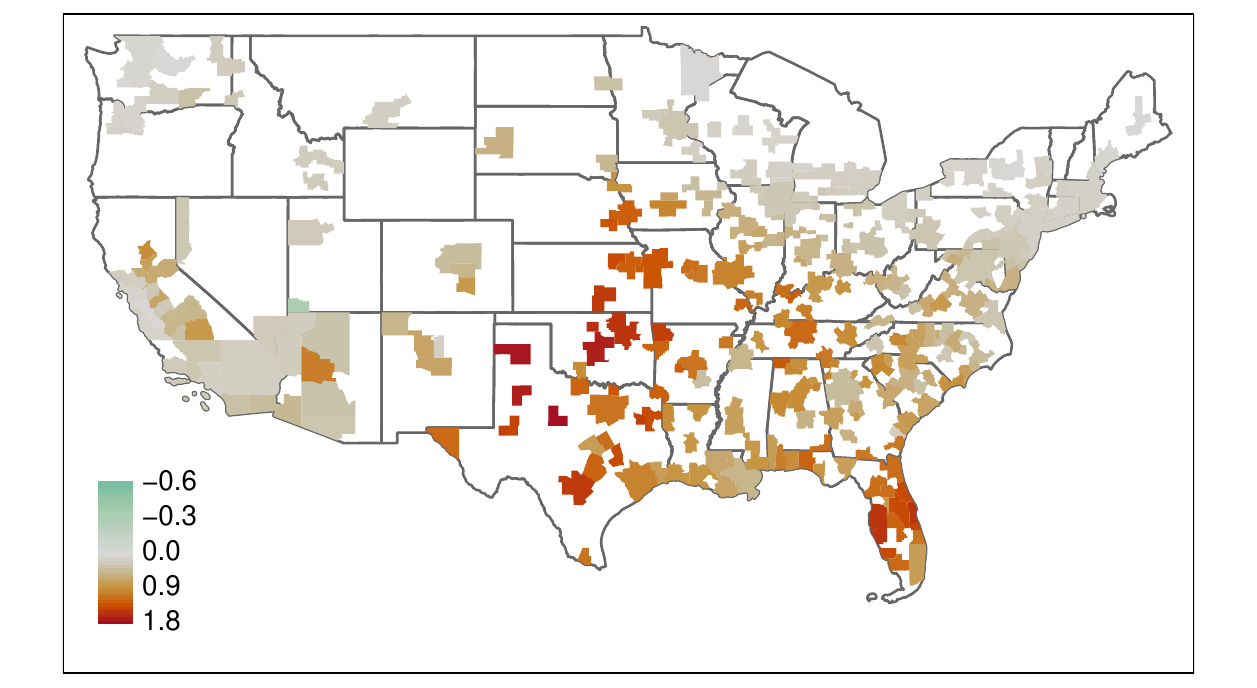}
		\caption{Under SSP5}
		\label{fig:ssp5}
    \end{subfigure}
    \caption{Projected change in experienced segregation under different climate scenarios    \label{fig:ssp}}
    \justifying \noindent \footnotesize{Projected change in the number of per capita encounters of non-White and White individuals in 2050 compared to the reference period from 1987 to 2017. Panel~\ref{fig:ssp3} refers to a best-case climate change scenario with a temperature rise of well below $2$\textdegree C by 2100, Panel~\ref{fig:ssp5} to a worst-case scenario with a temperature rise of $4.7$ to $5.1$\textdegree C.} 
\end{figure}

With progressing climate change, exposure to higher temperatures will increase. We aim to assess the potential net effect of such climatic changes on isolation. To this end, we conduct a simple \textit{ceteris paribus} projection of the change in the number of encounters a {non-\textit{White}} person today would have with a {\textit{White}} person if she lived in a future climate. We proceed as follows: First, we calculate the weekly difference in the number of days per temperature bin between the two climate scenarios in 2050 and the reference period from 1987 to 2017. Second, we multiply the temperature differences with the estimated coefficients $\hat{\beta}$ from Eq.~\ref{eq:estimate} and the total volume of visits. To derive the individual-level change in the number of encounters, we divide the quantities by the number of registered devices (see Fig.~\ref{fig:ssp}).\footnote{To obtain the population-wide change in the number of encounters per MSA, we alternatively multiply the calculated quantities with the ratio of MSA population size over the number of devices recorded by Safegraph (see Appendix Table~\ref{tab:clim_proj}).}

Importantly, this back-of-the-envelope projection neglects potential changes in the population composition and adaptation. Thus, it only serves as a first-order approximation of potential racial isolation under future climate change. We combine the temperature-isolation relationship estimated in Fig.~\ref{fig:main} with two possible climate scenarios from the Sixth Assessment Report of the Intergovernmental Panel on Climate Change (IPCC). A best-case scenario represents the implementation of strict climate policy that ensures a global temperature increase well below $2$\textdegree C by 2100 (SSP1/RCP2.6), and a worst-case scenario represents the absence of any emission reductions leading to a temperature rise of $4.7$ to $5.1$\textdegree C (SSP5/RCP8.5). We consider the change in between-group encounters in both scenarios compared to the reference period 1987-2017. 

Fig.~\ref{fig:ssp} shows that the encounters decrease in many MSAs in both scenarios in 2050. However, the number of MSAs that experience increasing isolation and the magnitude of the projected changes is larger in the worst-case scenario. Only one MSA in Utah exhibits a notable decrease in isolation. However, this effect decreases in magnitude from scenario SSP1/RCP2.6 to SSP5/RCP8.5, suggesting that the former scenario leads to less isolation than the latter for all MSAs despite substantial geographic heterogeneity. This suggests that strong climate policy has the co-benefit of reducing temperature-induced racial isolation under global warming. In Appendix Table~\ref{tab:clim_proj}, we also approximate the total number of foregone encounters that could accrue in the most affected MSAs due to a changing climate in 2050. In Dallas, for instance, global warming could eliminate up to about eight million between-group encounters compared to the annual average number of encounters in the reference period.

%% file: conclusion.tex
\hypertarget{conclusion}{%
\section{Conclusion}\label{conclusion}}

We find that exposure to daily maximum temperature above 25\dg C has a robust positive effect on an index of isolation that estimates the extent of segregation based on the probability of inter-racial encounters. Therefore, we conclude that heat positively affects racial segregation. The effect is particularly strong at places for leisure activities and for individuals living in lower-income areas. Across the temperature distribution, we find a U-shaped relationship, though effects of low temperature are mostly smaller in magnitude and less robust. We project that the net effect of climate change until the middle of the century is segregation-increasing in most MSAs. Compared to unabated warming, strict climate policy will consistently lead to less isolation across the US.

The existence of heat-induced segregation in daily life is hard to grasp quantitatively and our study provides the first evidence attesting their existence based on large-scale foot traffic data. Although the estimated effects remain moderate in absolute terms, they may indicate broader behavioral adjustments to temperatures that affect the interracial exchange and are undocumented thus far. With these insights, our study calls for further research. While we shed light on another channel through which global warming might affect social welfare, only future research can clarify just \textit{how} this effect materializes.

Further research is needed that provides explanations for racial differences in visiting outdoor and indoor leisure places during heat periods. Our analysis hints at income-related effect differences, and other studies confirm that poverty status is an important factor in explaining segregation \citep{Wang2018}. However, only a further analysis can elucidate the specific drivers of heat-caused isolation which may include racial differences in access to different modes of transportation or the distance of residency to the POI, for instance. In addition, future research could investigate potential segregation dynamics within the two groups of {\textit{White}} and {non-\textit{White}} individuals that we observe. We hypothesize that there are significant differences in the average socialization, living conditions, and domains of activity of African Americans, Hispanics, Asian Americans, Native Americans, and other groups that we ignore in our simplified two-group consideration.

%% file: appendix.tex
\hypertarget{appendix}{%
\section*{Appendix}\label{appendix}}

\section*{\large{Appendix Figures}}

\begin{figure}
	\centering
		\includegraphics[width = 0.8\linewidth]{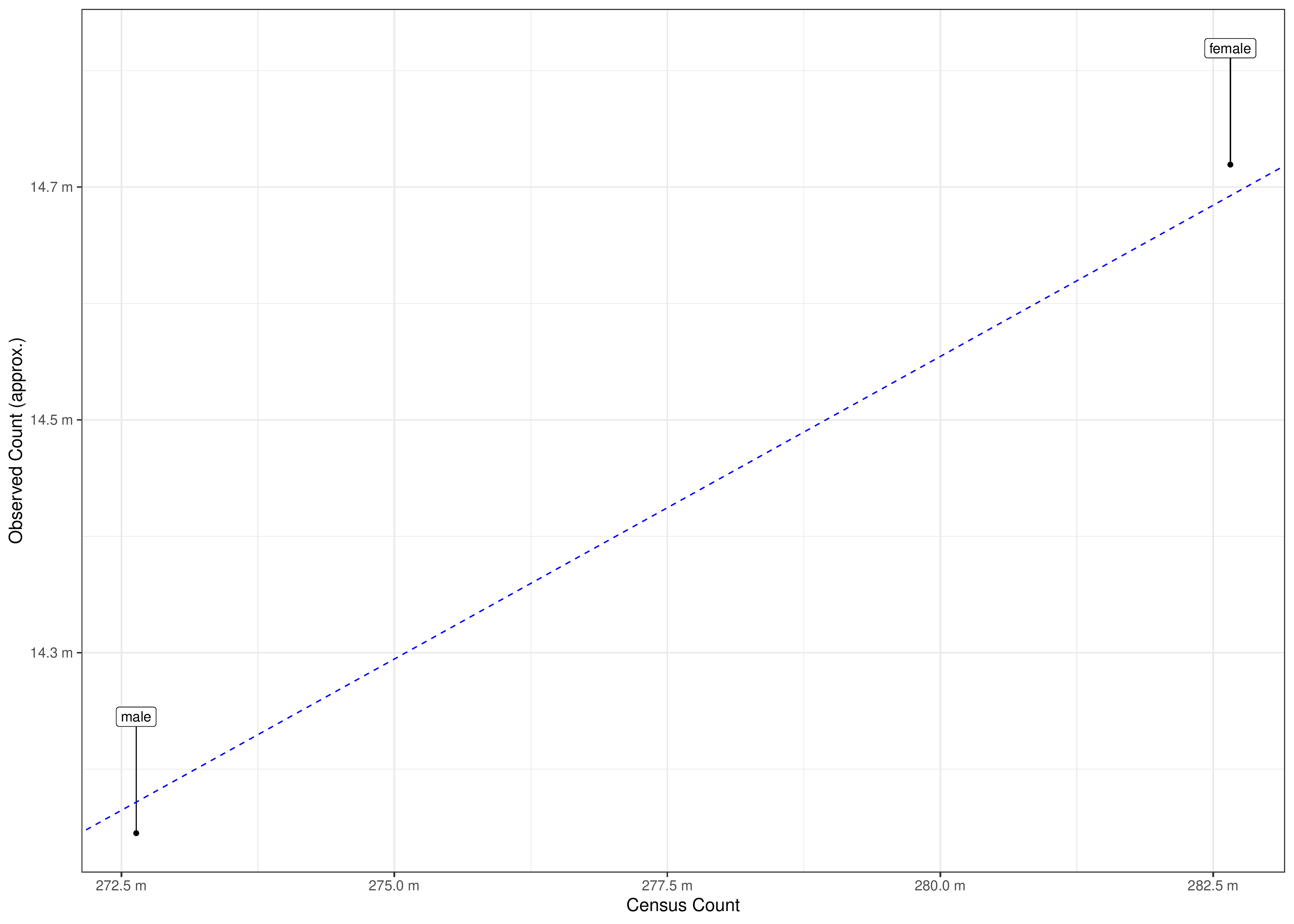}
		\caption{Representativeness of SafeGraph data in terms of sex \label{fig:app1}}
	\vspace{10pt}
	\justifying \noindent \footnotesize{The plot shows how the sex distribution of the SafeGraph users (y-axis) relates to the sex distribution recorded by the census (x-axis). Both axes count the number of individuals in million (m). For representativeness the counts should line up along the diagonal line in blue.}
\end{figure}

\begin{figure}
	\centering
		\includegraphics[width = 0.8\linewidth]{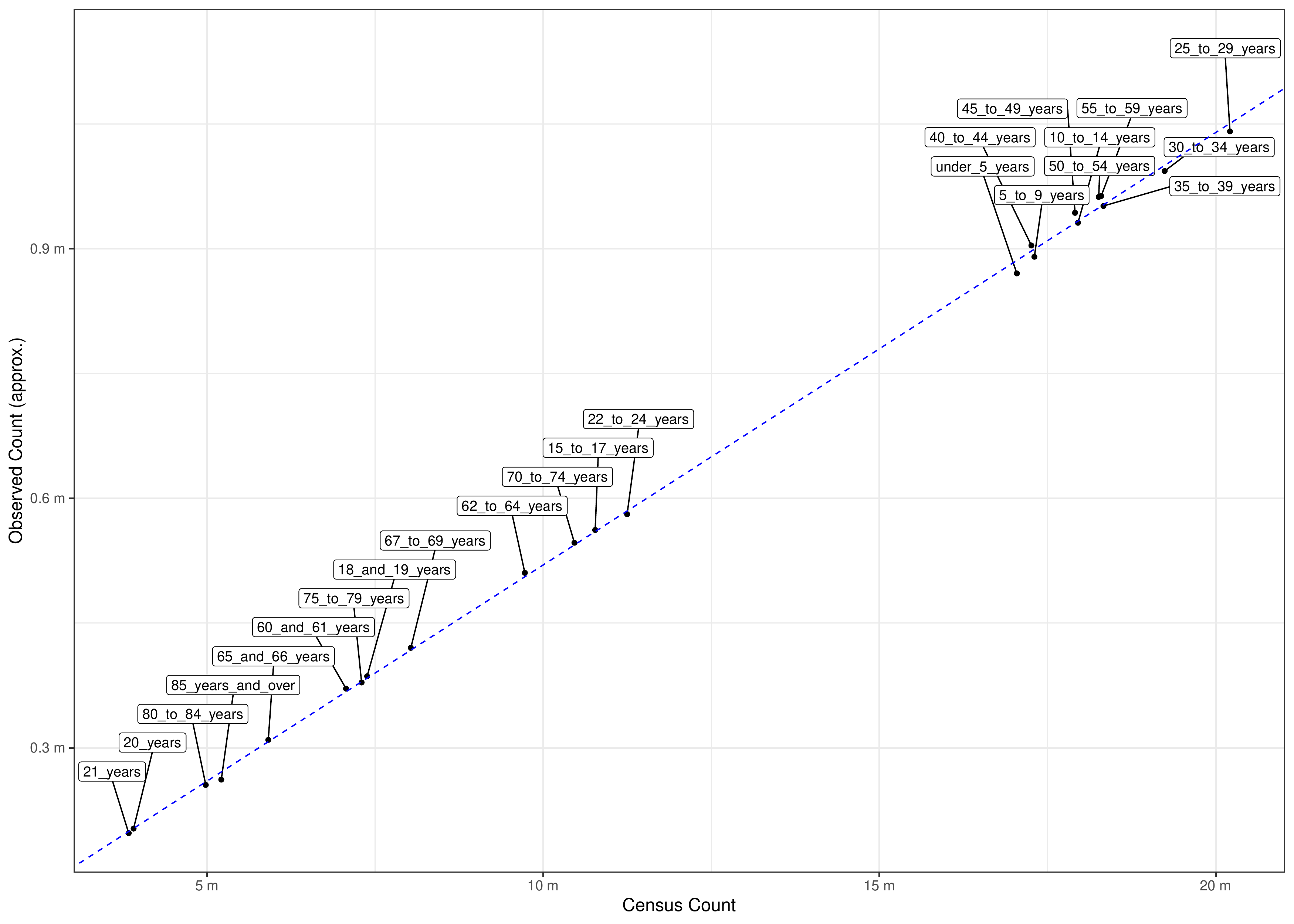}
		\caption{Representativeness of SafeGraph data in terms of age\label{fig:app2}}
	\vspace{10pt}
	\justifying \noindent \footnotesize{The plot shows how the age distribution of the SafeGraph users (y-axis) relates to the age distribution recorded by the census (x-axis). Both axes count the number of individuals in million (m). For representativeness the counts should line up along the diagonal line in blue.}
\end{figure}

\begin{figure}
	\centering
		\includegraphics[width = 0.8\linewidth]{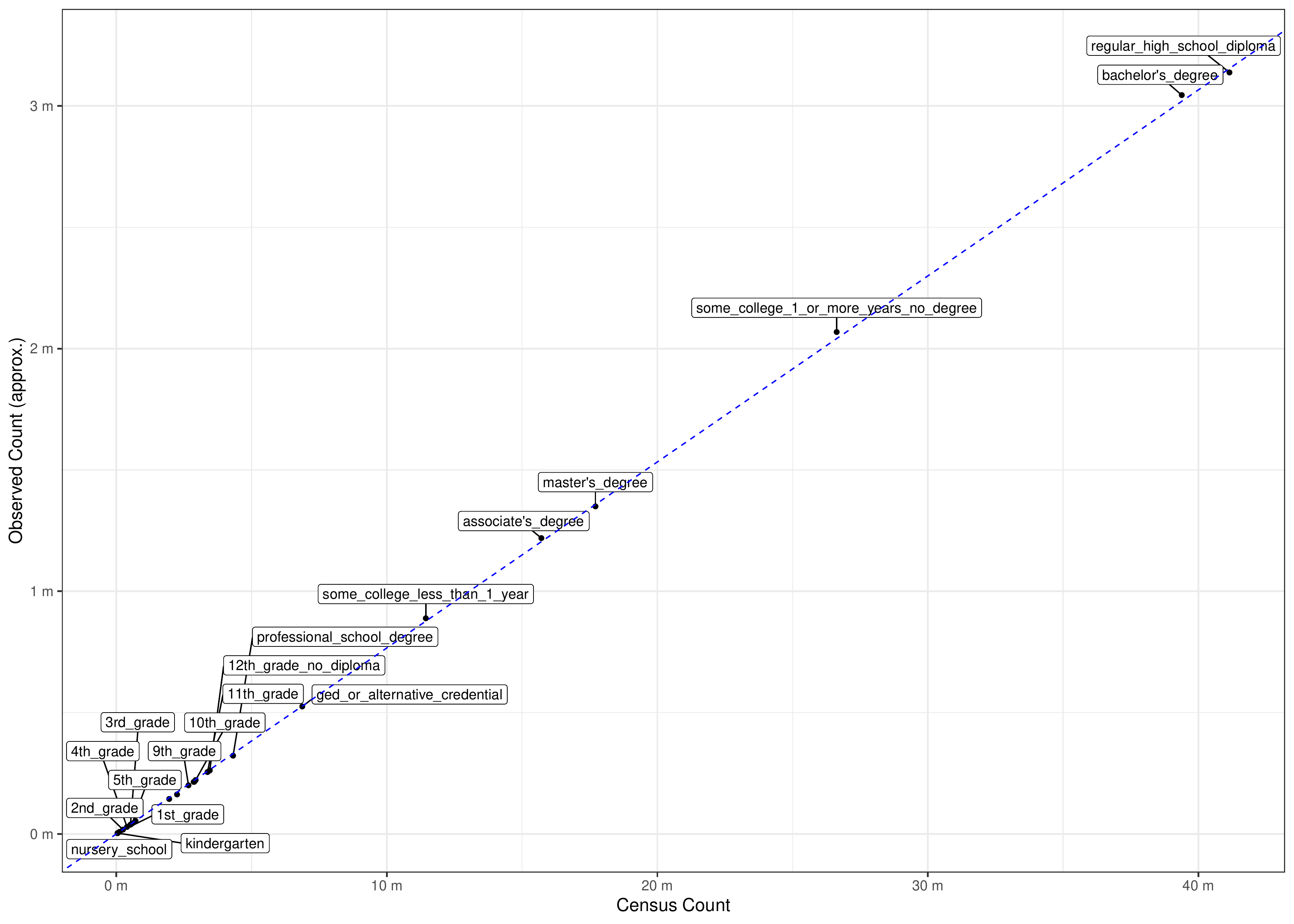}
		\caption{Representativeness of SafeGraph data in terms of education \label{fig:app3}}
	\vspace{10pt}
	\justifying \noindent \footnotesize{The plot shows how the education distribution of the SafeGraph users (y-axis) relates to the education distribution recorded by the census (x-axis). Both axes count the number of individuals in million (m). For representativeness the counts should line up along the diagonal line in blue.}
\end{figure}

\begin{figure}
	\centering
		\includegraphics[width = 0.8\linewidth]{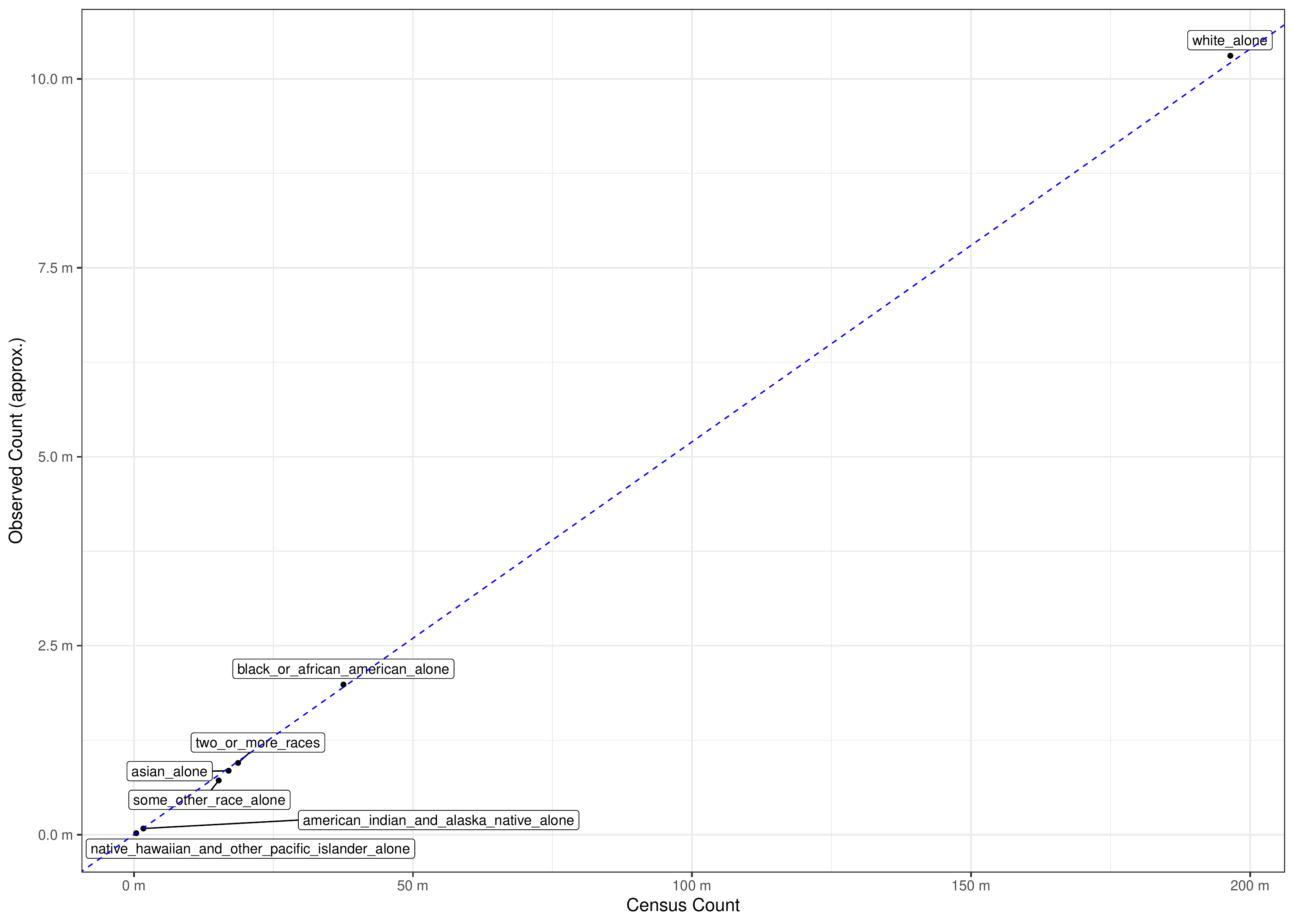}
		\caption{Representativeness of SafeGraph data in terms of race\label{fig:app4}}
	\vspace{10pt}
	\justifying \noindent \footnotesize{The plot shows how the race distribution of the SafeGraph users (y-axis) relates to the race distribution recorded by the census (x-axis). Both axes count the number of individuals in million (m). For representativeness the counts should line up along the diagonal line in blue.}
\end{figure}

\begin{figure}
	\centering
		\includegraphics[width = 0.8\linewidth]{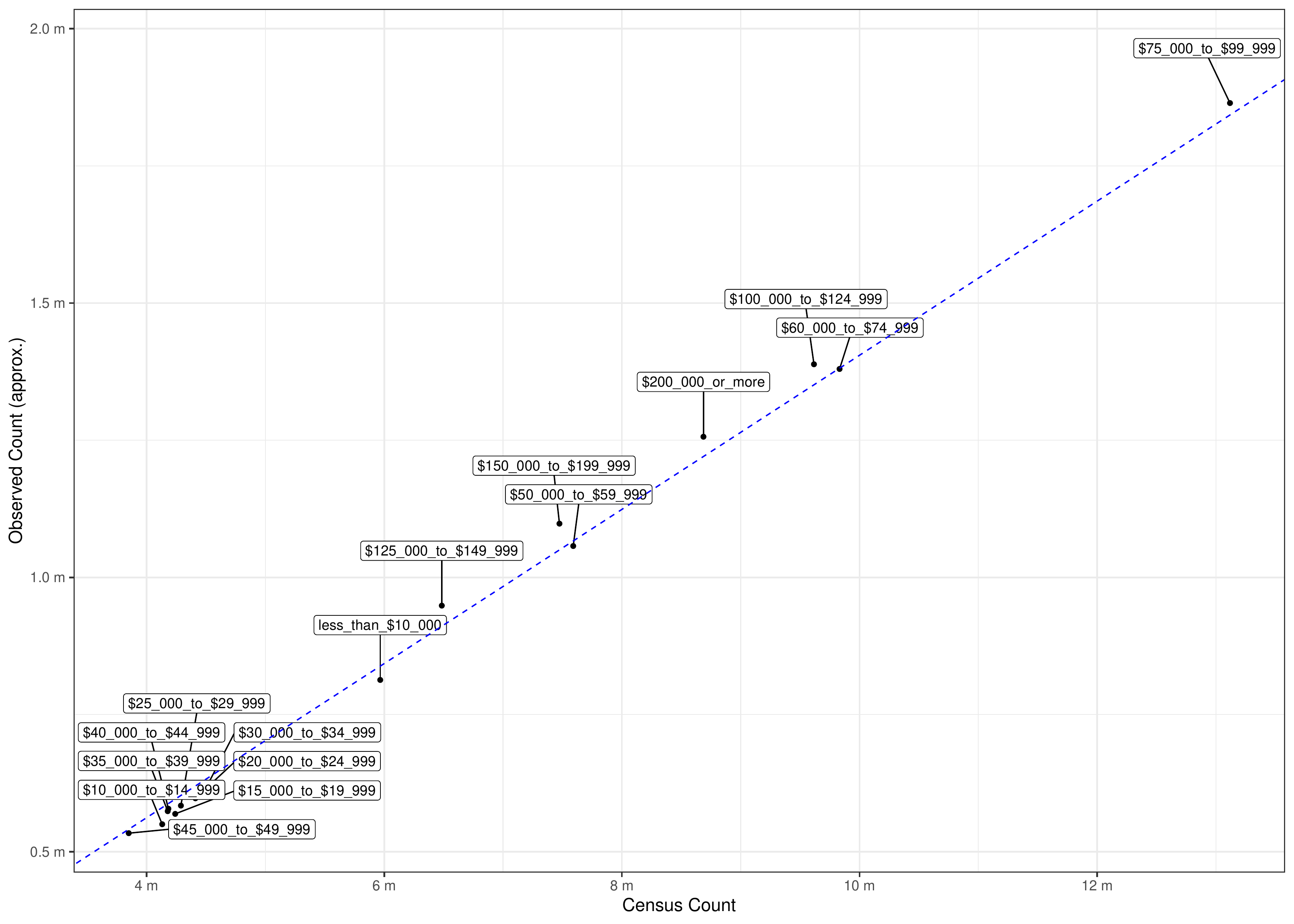}
		\caption{Representativeness of SafeGraph data in terms of income \label{fig:app5}}
	\vspace{10pt}
	\justifying \noindent \footnotesize{The plot shows how the income distribution of the SafeGraph users (y-axis) relates to the income distribution recorded by the census (x-axis). Both axes count the number of individuals in million (m). For representativeness the counts should line up along the diagonal line in blue.}
\end{figure}

\begin{figure}
	{\centering
		\includegraphics[width = 0.8\linewidth]{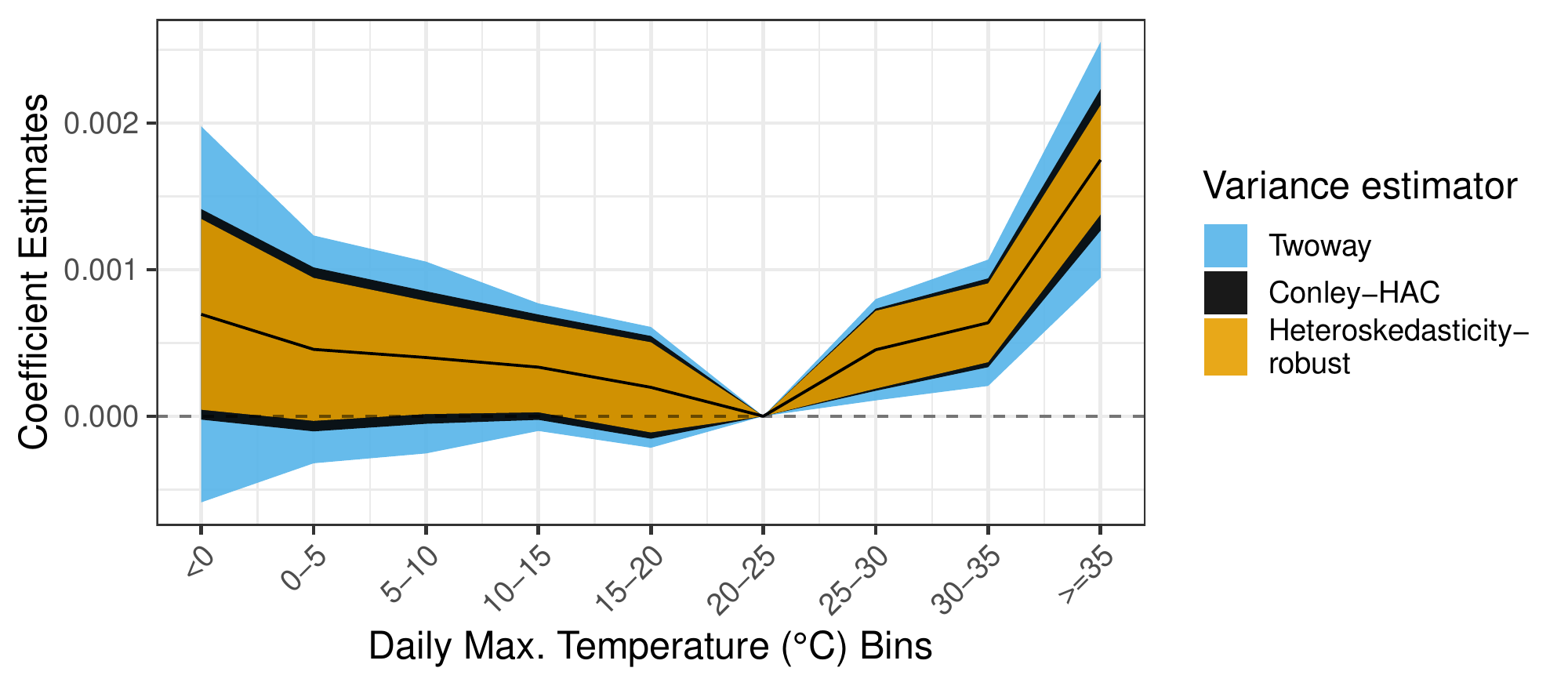}
		\caption{Comparison to robust and twoway clustered standard errors}
		\label{fig:se}}
	\vspace{10pt}
	\justifying \noindent \footnotesize{The figure shows the coefficients estimated for each of the daily maximum temperature bins excluding the 20\textdegree C to 25\textdegree C reference bin based on Eq.~\ref{eq:estimate} and weighting by MSA population size. The colored areas depict asymptotic 95\% confidence intervals based on heteroskedasticity-robust standard errors in orange, Conley-HAC standard errors in black, and twoway clustered standard errors in blue.}
\end{figure}

\begin{figure}
	{\centering
		\includegraphics[width = 0.8\linewidth]{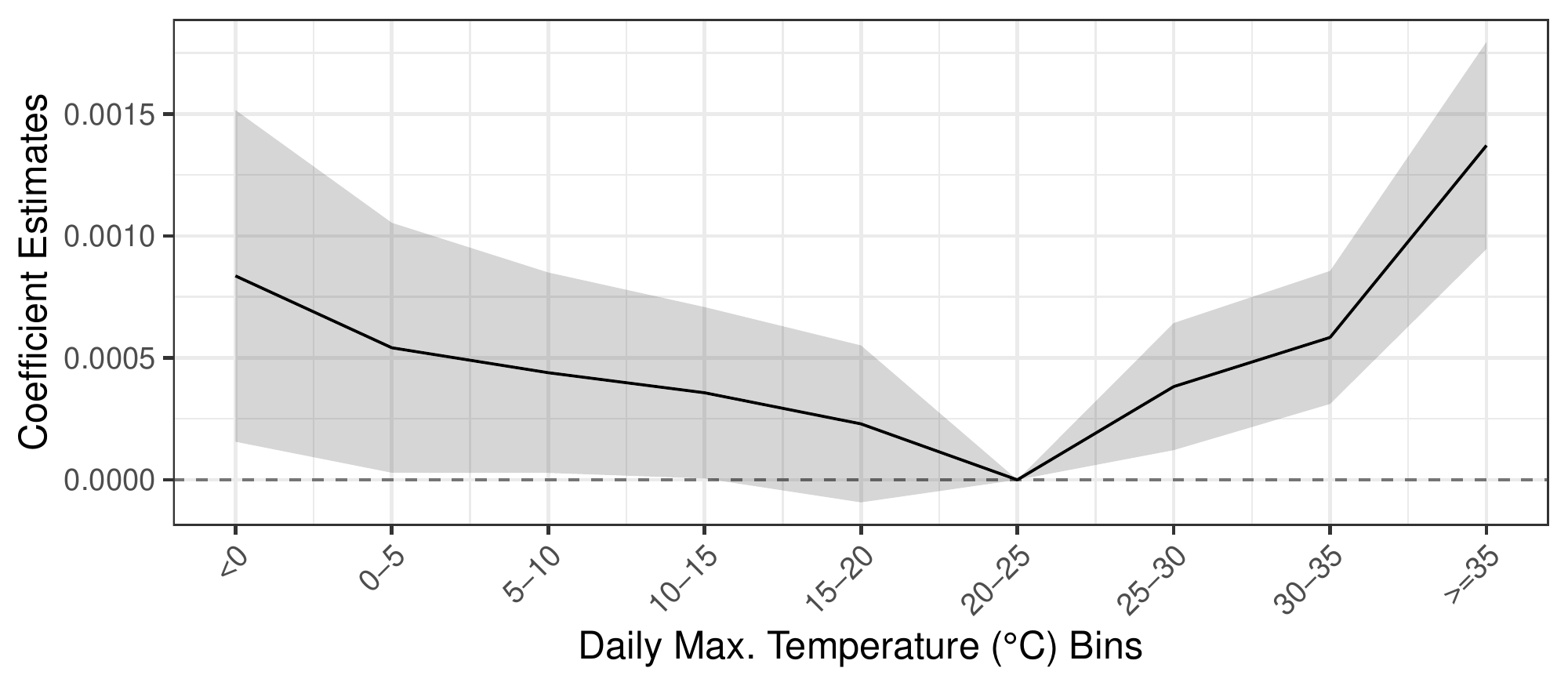}
		\caption{Sample based on urban areas}
		\label{fig:UA}}
	\vspace{10pt}
\justifying \noindent \footnotesize{The figure shows the coefficients estimated for each of the daily maximum temperature bins excluding the 20\textdegree C to 25\textdegree C reference bin based on Eq.~\ref{eq:estimate}. All variables are derived on the urban area (UA) level and observations are weighted by UA population size. The shaded areas depict asymptotic 95\% confidence intervals based on Conley-HAC standard errors.}
\end{figure}

\begin{figure}
	\centering
		\includegraphics[width = 0.8\linewidth]{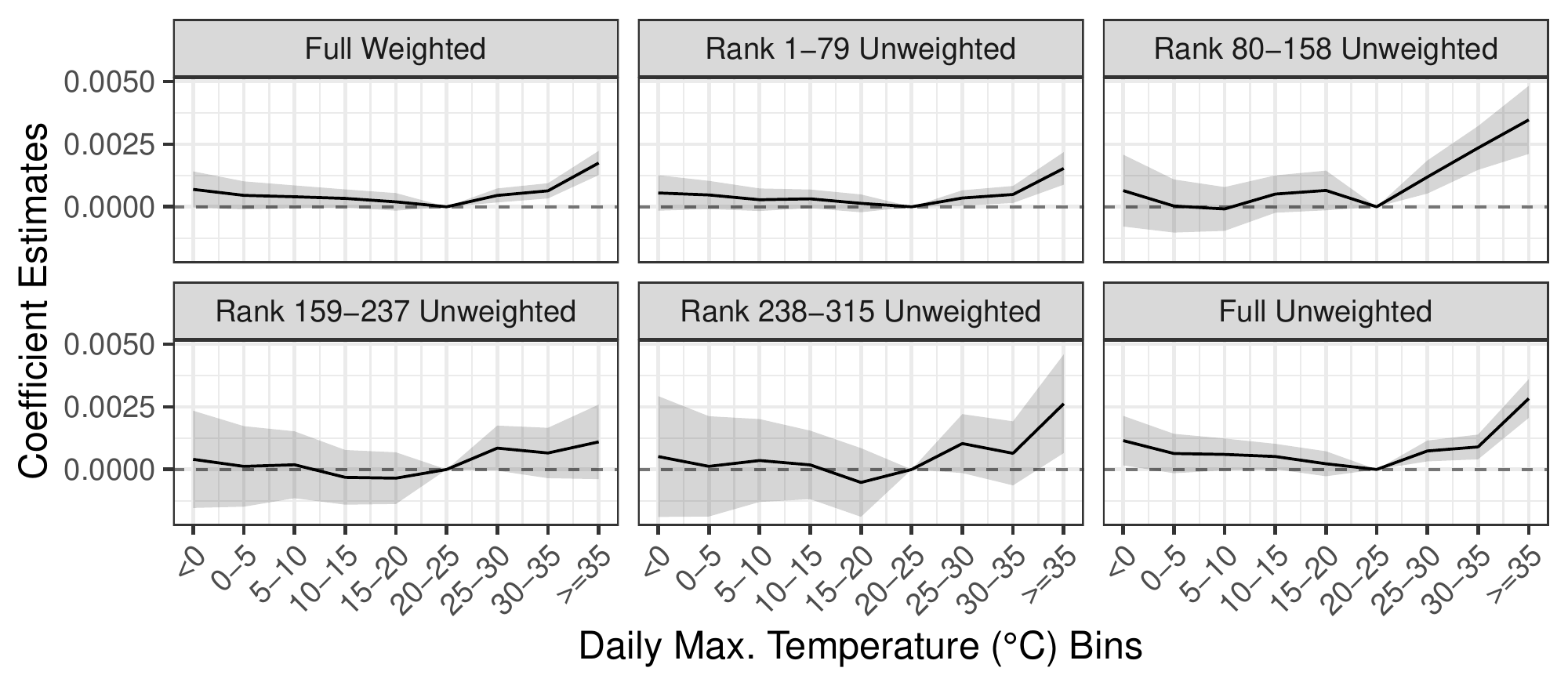}
		\caption{Estimations based on subsamples with different population size\label{fig:popsplit}}
  \justifying \noindent \footnotesize{Coefficients estimated for each of the daily maximum temperature bins excluding the 20\textdegree C to 25\textdegree C reference bin based on Eq.~\ref{eq:estimate}. We consider six subsamples based on MSA population size as indicated by the respective subheadings. The subheadings also indicate whether observations are weighted by MSA population size. The shaded areas depict asymptotic 95\% confidence intervals based on Conley-HAC standard errors.}
\end{figure}
\FloatBarrier

\section*{\large{Appendix Tables}}

\begin{table}
\begin{tabular}{lrr}
    \toprule MSA & \multicolumn{2}{c}{\begin{tabular}{c}Present-2050 total change \\ in encounters (in millions)\end{tabular}} \\
     & {under SSP5/RCP8.5} & {under SSP1/RCP2.6} \\
    \midrule Dallas, TX & 8.0 & 4.9 \\
     Houston, TX & 6.9 & 4.8 \\
     Tampa, FL & 4.7 & 3.5 \\
     Miami, FL & 4.7 & 2.6 \\
     San Antonio, TX & 3.7 & 2.5 \\
     Orlando, FL & 3.5 & 2.7 \\
     Kansas City, MO-KS & 2.8 & 1.9 \\
     St. Louis, MO-IL & 2.7 & 1.5 \\
     Atlanta, GA & 2.7 & 1.4 \\
     Chicago, IL-IN-WI & 2.7 & 1.4 \\
    \bottomrule
\end{tabular}
\caption{Top ten cities with decreasing encounters}
\centering \noindent \footnotesize{Projected decreases in the total number of encounters in millions per city in 2050 under climate scenarios SSP1/RCP2.6 and SSP5/RCP8.5.}
\label{tab:clim_proj}
\end{table}

%% file: jablib.bib
@article{marz2015intergenerational,
  title={Intergenerational Social Mobility in the {U}nited {S}tates: A Multivariate Analysis Using {B}ayesian Distributional Regression},
  author={März, Alexander and Klein, Nadja and Kneib, Thomas and Mußhoff, Oliver},
  year={2022},
  month = {11},
  journal={Working paper. Available on request.}
}

@article{lange_buechner_2021,
	title = {ISIMIP3b Bias-adjusted Atmospheric Climate Input 
 data (v1.1)},
	journal = {ISIMIP Repository},
	author = {Lange, Stefan and Büchner, Matthias},
	year = {2021},
 month = {05},
 doi = {https://doi.org/10.48364/ISIMIP.842396.1}
}

@Article{Cutler1997,
	author   = {Cutler, David M. and Glaeser, Edward L.},
	journal  = {The Quarterly Journal of Economics},
	title    = {Are Ghettos Good or Bad?},
	year     = {1997},
	issn     = {0033-5533},
	month    = {08},
	number   = {3},
	pages    = {827-872},
	volume   = {112},
	doi      = {10.1162/003355397555361},
	eprint   = {https://academic.oup.com/qje/article-pdf/112/3/827/5431883/112-3-827.pdf},
	url      = {https://doi.org/10.1162/003355397555361},
}

@Article{Burke2018,
	author          = {Burke, Marshall and Gonz{\'{a}}lez, Felipe and Baylis, Patrick and Heft-Neal, Sam and Baysan, Ceren and Basu, Sanjay and Hsiang, Solomon},
	journal         = {Nature Climate Change},
	title           = {Higher Temperatures Increase Suicide Rates in the {{United States}} and {Mexico}},
	year            = {2018},
	pages = {723-729},
	volume = {8},
	xdoi            = {10.1038/s41558-018-0222-x},
	xissn           = {17586798},
}

@Article{Deschenes2007,
Author = {Deschênes, Olivier and Greenstone, Michael},
Title = {The Economic Impacts of Climate Change: Evidence from Agricultural Output and Random Fluctuations in Weather},
Journal = {American Economic Review},
Volume = {97},
Number = {1},
Year = {2007},
Month = {March},
Pages = {354-385},
DOI = {10.1257/aer.97.1.354},
URL = {https://www.aeaweb.org/articles?id=10.1257/aer.97.1.354}}

@Article{Hsiang2016,
  author   = {Hsiang, Solomon},
  journal  = {Annual Review of Resource Economics},
  title    = {Climate Econometrics},
  year     = {2016},
  number   = {1},
  pages    = {43-75},
  volume   = {8},
  doi      = {10.1146/annurev-resource-100815-095343},
  eprint   = {https://doi.org/10.1146/annurev-resource-100815-095343},
  url      = {https://doi.org/10.1146/annurev-resource-100815-095343},
}

@Article{Park2021,
	author          = {Park, Jisung and Pankratz, Nora M. C. and Behrer, A.},
	journal         = {IZA DP},
	volume   = {No. 14560},
	title           = {Temperature, Workplace Safety, and Labor Market Inequality},
	year            = {2021},
	xdoi            = {10.2139/ssrn.3892588},
}

@Misc{PRISMClimateGroup2022,
	author          = {{PRISM Climate Group}},
	howpublished         = {\url{http://prism.oregonstate.edu}},
	title           = {{PRISM Climate Group}, {Oregon State University}},
	year            = {2022},
	note = {Accessed December 4, 2022}
}

@Article{Wang2018,
  author   = {Qi Wang and Nolan Edward Phillips and Mario L. Small and Robert J. Sampson},
  journal  = {Proceedings of the National Academy of Sciences},
  title    = {Urban mobility and neighborhood isolation in America’s 50 largest cities},
  year     = {2018},
  number   = {30},
  pages    = {7735-7740},
  volume   = {115},
  doi      = {10.1073/pnas.1802537115},
  eprint   = {https://www.pnas.org/doi/pdf/10.1073/pnas.1802537115},
  url      = {https://www.pnas.org/doi/abs/10.1073/pnas.1802537115},
}

@Article{Zivin2014,
 ISSN = {0734306X, 15375307},
URL = {http://www.jstor.org/stable/10.1086/671766},
author = {Joshua Graff Zivin and Matthew Neidell},
journal = {Journal of Labor Economics},
number = {1},
pages = {1--26},
publisher = {[The University of Chicago Press, Society of Labor Economists, NORC at the University of Chicago]},
title = {Temperature and the Allocation of Time: Implications for Climate Change},
urldate = {2022-12-04},
volume = {32},
year = {2014}
}

@Article{Athey2021,
	author   = {Susan Athey and Billy Ferguson and Matthew Gentzkow and Tobias Schmidt},
	journal  = {Proceedings of the National Academy of Sciences},
	title    = {Estimating experienced racial segregation in {US} cities using large-scale {GPS} data},
	year     = {2021},
	number   = {46},
	pages    = {e2026160118},
	volume   = {118},
	doi      = {10.1073/pnas.2026160118},
	eprint   = {https://www.pnas.org/doi/pdf/10.1073/pnas.2026160118},
	url      = {https://www.pnas.org/doi/abs/10.1073/pnas.2026160118},
}

@Article{Echenique2007,
	author   = {Echenique, Federico and Fryer, Roland G., Jr.},
	journal  = {The Quarterly Journal of Economics},
	title    = {A Measure of Segregation Based on Social Interactions},
	year     = {2007},
	issn     = {0033-5533},
	month    = {05},
	number   = {2},
	pages    = {441-485},
	volume   = {122},
	doi      = {10.1162/qjec.122.2.441},
	eprint   = {https://academic.oup.com/qje/article-pdf/122/2/441/5470068/122-2-441.pdf},
	url      = {https://doi.org/10.1162/qjec.122.2.441},
}

@Article{Davis2019,
	author   = {Davis, Donald R. and Dingel, Jonathan I. and Monras, Joan and Morales, Eduardo},
	journal  = {Journal of Political Economy},
	title    = {How Segregated Is Urban Consumption?},
	year     = {2019},
	number   = {4},
	pages    = {1684-1738},
	volume   = {127},
	doi      = {10.1086/701680},
	eprint   = {https://doi.org/10.1086/701680},
	url      = {https://doi.org/10.1086/701680},
}

@Article{Gentzkow2011,
	author   = {Gentzkow, Matthew and Shapiro, Jesse M.},
	journal  = {The Quarterly Journal of Economics},
	title    = {Ideological Segregation Online and Offline},
	year     = {2011},
	issn     = {0033-5533},
	month    = {11},
	number   = {4},
	pages    = {1799-1839},
	volume   = {126},
	doi      = {10.1093/qje/qjr044},
	eprint   = {https://academic.oup.com/qje/article-pdf/126/4/1799/17089890/qjr044.pdf},
	url      = {https://doi.org/10.1093/qje/qjr044},
}

@Article{Wong2011,
	author   = {Wong, David W. S. and Shaw, Shih-Lung},
	journal  = {Journal of Geographical Systems},
	title    = {Measuring segregation: an activity space approach},
	year     = {2011},
	issn     = {1435-5949},
	number   = {2},
	pages    = {127--145},
	volume   = {13},
	doi      = {10.1007/s10109-010-0112-x},
	refid    = {Wong2011},
	url      = {https://doi.org/10.1007/s10109-010-0112-x},
}

@Misc{Falcone2016,
	author   = {James A Falcone},
	title    = {{U.S.} block-level population density rasters for 1990, 2000, and 2010: {U.S. Geological Survey} data release},
	year     = {2016},
	howpublished      = {\url{http://dx.doi.org/10.5066/F74J0C6M}},
	note = {Accessed December 4, 2022}
}

@Article{Dell2014,
	author  = {Dell, Melissa and Jones, Benjamin F. and Olken, Benjamin A.},
	journal = {Journal of Economic Literature},
	title   = {What Do We Learn from the Weather? {The} New Climate-Economy Literature},
	year    = {2014},
	month   = {September},
	number  = {3},
	pages   = {740-98},
	volume  = {52},
	doi     = {10.1257/jel.52.3.740},
	url     = {https://www.aeaweb.org/articles?id=10.1257/jel.52.3.740},
}

@Book{Wooldridge2010,
	author    = {Jeffrey M. Wooldridge},
	publisher = {The MIT Press},
	title     = {Econometric Analysis of Cross Section and Panel Data, 2nd ed},
	year      = {2010},
	isbn      = {9780262232586},
	url       = {http://www.jstor.org/stable/j.ctt5hhcfr},
	urldate   = {2022-10-18},
}

@TechReport{Chen2020,
	author      = {Chen, M. Keith and Pope, Devin G},
	institution = {National Bureau of Economic Research},
	title       = {Geographic Mobility in {America}: Evidence from Cell Phone Data},
	year        = {2020},
	month       = {May},
	number      = {27072},
	type        = {Working Paper},
	doi         = {10.3386/w27072},
	series      = {Working Paper Series},
	url         = {http://www.nber.org/papers/w27072},
}

@Article{Prestby2020,
	author   = {Timothy Prestby and Joseph App and Yuhao Kang and Song Gao},
	journal  = {Environment and Planning A: Economy and Space},
	title    = {Understanding neighborhood isolation through spatial interaction network analysis using location big data},
	year     = {2020},
	number   = {6},
	pages    = {1027-1031},
	volume   = {52},
	doi      = {10.1177/0308518X19891911},
	eprint   = {https://doi.org/10.1177/0308518X19891911},
	url      = {https://doi.org/10.1177/0308518X19891911},
}

@Article{Jay2022,
	author   = {Jonathan Jay and Felicia Heykoop and Linda Hwang and Alexa Courtepatte and Jorrit {de Jong} and Michelle Kondo},
	journal  = {Landscape and Urban Planning},
	title    = {Use of smartphone mobility data to analyze city park visits during the {COVID}-19 pandemic},
	year     = {2022},
	issn     = {0169-2046},
	pages    = {104554},
	volume   = {228},
	doi      = {https://doi.org/10.1016/j.landurbplan.2022.104554},
	url      = {https://www.sciencedirect.com/science/article/pii/S0169204622002031},
}

@Article{Ma2021,
	author   = {Ma, Yiqun and Pei, Sen and Shaman, Jeffrey and Dubrow, Robert and Chen, Kai},
	journal  = {Nature Communications},
	title    = {Role of meteorological factors in the transmission of {SARS-CoV-2} in the {United States}},
	year     = {2021},
	issn     = {2041-1723},
	number   = {1},
	pages    = {3602},
	volume   = {12},
	doi      = {10.1038/s41467-021-23866-7},
	refid    = {Ma2021},
	url      = {https://doi.org/10.1038/s41467-021-23866-7},
}

@Article{Li2022,
	author   = {Xiao Li and Xiao Huang and Dongying Li and Yang Xu},
	journal  = {Sustainable Cities and Society},
	title    = {Aggravated social segregation during the {COVID}-19 pandemic: Evidence from crowdsourced mobility data in twelve most populated {U.S.} metropolitan areas},
	year     = {2022},
	issn     = {2210-6707},
	pages    = {103869},
	volume   = {81},
	doi      = {https://doi.org/10.1016/j.scs.2022.103869},
	url      = {https://www.sciencedirect.com/science/article/pii/S2210670722001962},
}

@Article{Allcott2020,
	author   = {Hunt Allcott and Levi Boxell and Jacob Conway and Matthew Gentzkow and Michael Thaler and David Yang},
	journal  = {Journal of Public Economics},
	title    = {Polarization and public health: Partisan differences in social distancing during the coronavirus pandemic},
	year     = {2020},
	issn     = {0047-2727},
	pages    = {104254},
	volume   = {191},
	doi      = {https://doi.org/10.1016/j.jpubeco.2020.104254},
	url      = {https://www.sciencedirect.com/science/article/pii/S0047272720301183},
}

@Article{Benzell2020,
	author   = {Seth G. Benzell and Avinash Collis and Christos Nicolaides},
	journal  = {Proceedings of the National Academy of Sciences},
	title    = {Rationing social contact during the {COVID}-19 pandemic: Transmission risk and social benefits of {US} locations},
	year     = {2020},
	number   = {26},
	pages    = {14642-14644},
	volume   = {117},
	doi      = {10.1073/pnas.2008025117},
	eprint   = {https://www.pnas.org/doi/pdf/10.1073/pnas.2008025117},
	url      = {https://www.pnas.org/doi/abs/10.1073/pnas.2008025117},
}

@Article{Chang2021,
	author   = {Chang, Serina and Pierson, Emma and Koh, Pang Wei and Gerardin, Jaline and Redbird, Beth and Grusky, David and Leskovec, Jure},
	journal  = {Nature},
	title    = {Mobility network models of {COVID}-19 explain inequities and inform reopening},
	year     = {2021},
	issn     = {1476-4687},
	number   = {7840},
	pages    = {82--87},
	volume   = {589},
	doi      = {10.1038/s41586-020-2923-3},
	refid    = {Chang2021},
	url      = {https://doi.org/10.1038/s41586-020-2923-3},
}

@Article{Chetty2014,
	author  = {Chetty, Raj and Hendren, Nathaniel and Kline, Patrick and Saez, Emmanuel and Turner, Nicholas},
	journal = {American Economic Review},
	title   = {Is the {{United States}} Still a Land of Opportunity? {Recent} Trends in Intergenerational Mobility},
	year    = {2014},
	month   = {May},
	number  = {5},
	pages   = {141-47},
	volume  = {104},
	doi     = {10.1257/aer.104.5.141},
	url     = {https://www.aeaweb.org/articles?id=10.1257/aer.104.5.141},
}

@Article{Chetty2018b,
	author   = {Chetty, Raj and Hendren, Nathaniel},
	journal  = {The Quarterly Journal of Economics},
	title    = {The Impacts of Neighborhoods on Intergenerational Mobility {I}: Childhood Exposure Effects},
	year     = {2018},
	issn     = {0033-5533},
	month    = {02},
	number   = {3},
	pages    = {1107-1162},
	volume   = {133},
	doi      = {10.1093/qje/qjy007},
	eprint   = {https://academic.oup.com/qje/article-pdf/133/3/1107/25705047/qjy007.pdf},
	url      = {https://doi.org/10.1093/qje/qjy007},
}

@Article{Massey1990,
	author   = {Massey, Douglas S.},
	journal  = {American Journal of Sociology},
	title    = {{American} Apartheid: Segregation and the Making of the Underclass},
	year     = {1990},
	number   = {2},
	pages    = {329-357},
	volume   = {96},
	doi      = {10.1086/229532},
	eprint   = {https://doi.org/10.1086/229532},
	url      = {https://doi.org/10.1086/229532},
}

@Article{Kain1968,
	author   = {Kain, John F.},
	journal  = {The Quarterly Journal of Economics},
	title    = {Housing Segregation, {Negro} Employment, and Metropolitan Decentralization},
	year     = {1968},
	issn     = {0033-5533},
	month    = {05},
	number   = {2},
	pages    = {175-197},
	volume   = {82},
	doi      = {10.2307/1885893},
	eprint   = {https://academic.oup.com/qje/article-pdf/82/2/175/5350689/82-2-175.pdf},
	url      = {https://doi.org/10.2307/1885893},
}

@Article{Collins1999,
	author    = {Chiquita A. Collins and David R. Williams},
	journal   = {Sociological Forum},
	title     = {Segregation and Mortality: The Deadly Effects of Racism?},
	year      = {1999},
	issn      = {08848971, 15737861},
	number    = {3},
	pages     = {495--523},
	volume    = {14},
	publisher = {[Wiley, Springer]},
	url       = {http://www.jstor.org/stable/684876},
	urldate   = {2022-10-25},
}

@Article{Almond2006,
	author  = {Almond, Douglas and Chay, Kenneth and Greenstone, Michael},
	journal = {MIT Department of Economics Working Paper},
	title   = {Civil Rights, the War on Poverty, and {Black-White} Convergence in Infant Mortality in the Rural {South} and {Mississippi}},
	year    = {2006},
	volume = {07},
	number = {04},
	month   = {12},
	doi     = {10.2139/ssrn.961021},
}

@Article{Krivo2009,
	author   = {Krivo, Lauren J. and Peterson, Ruth D. and Kuhl, Danielle C.},
	journal  = {American Journal of Sociology},
	title    = {Segregation, Racial Structure, and Neighborhood Violent Crime},
	year     = {2009},
	number   = {6},
	pages    = {1765-1802},
	volume   = {114},
	doi      = {10.1086/597285},
	eprint   = {https://doi.org/10.1086/597285},
	url      = {https://doi.org/10.1086/597285},
}

@Article{Light2019,
	author   = {Michael T. Light and Julia T. Thomas},
	journal  = {American Sociological Review},
	title    = {Segregation and Violence Reconsidered: Do {Whites} Benefit from Residential Segregation?},
	year     = {2019},
	number   = {4},
	pages    = {690-725},
	volume   = {84},
	doi      = {10.1177/0003122419858731},
	eprint   = {https://doi.org/10.1177/0003122419858731},
	url      = {https://doi.org/10.1177/0003122419858731},
}

@Article{Massey1995,
	author    = {Douglas S. Massey},
	journal   = {University of Pennsylvania Law Review},
	title     = {Getting Away with Murder: Segregation and Violent Crime in Urban {America}},
	year      = {1995},
	issn      = {00419907},
	number    = {5},
	pages     = {1203--1232},
	volume    = {143},
	publisher = {The University of Pennsylvania Law Review},
	url       = {http://www.jstor.org/stable/3312474},
	urldate   = {2022-10-25},
}

@Article{Sharkey2011,
	author   = {Sharkey, Patrick and Elwert, Felix},
	journal  = {American Journal of Sociology},
	title    = {The Legacy of Disadvantage: Multigenerational Neighborhood Effects on Cognitive Ability},
	year     = {2011},
	number   = {6},
	pages    = {1934-1981},
	volume   = {116},
	doi      = {10.1086/660009},
	eprint   = {https://doi.org/10.1086/660009},
	url      = {https://doi.org/10.1086/660009},
}

@Article{Massey1987,
	author   = {Massey, Douglas S. and Condran, Gretchen A. and Denton, Nancy A.},
	journal  = {Social Forces},
	title    = {The Effect of Residential Segregation on {Black} Social and Economic Well-Being},
	year     = {1987},
	issn     = {0037-7732},
	month    = {09},
	number   = {1},
	pages    = {29-56},
	volume   = {66},
	doi      = {10.1093/sf/66.1.29},
	eprint   = {https://academic.oup.com/sf/article-pdf/66/1/29/6515004/66-1-29.pdf},
	url      = {https://doi.org/10.1093/sf/66.1.29},
}

@TechReport{Cook2022,
	author      = {Cook, Cody and Currier, Lindsey and Glaeser, Edward L},
	institution = {National Bureau of Economic Research},
	title       = {Urban Mobility and the Experienced Isolation of Students and Adults},
	year        = {2022},
	month       = {January},
	number      = {29645},
	type        = {Working Paper},
	doi         = {10.3386/w29645},
	series      = {Working Paper Series},
	url         = {http://www.nber.org/papers/w29645},
}

@Article{Carleton2016,
	author   = {Tamma A. Carleton and Solomon M. Hsiang},
	journal  = {Science},
	title    = {Social and economic impacts of climate},
	year     = {2016},
	number   = {6304},
	pages    = {aad9837},
	volume   = {353},
	doi      = {10.1126/science.aad9837},
	eprint   = {https://www.science.org/doi/pdf/10.1126/science.aad9837},
	url      = {https://www.science.org/doi/abs/10.1126/science.aad9837},
}

@Misc{SafeGraph2022,
	author = {{SafeGraph}},
	note   = {Accessed Oct 21, 2022},
	title  = {{SafeGraph} Docs},
	year   = {2022},
	howpublished = {\url{https://docs.safegraph.com/docs}},
}

@InCollection{Ortiz2021,
	author    = {Ariel Ortiz-Bobea},
	booktitle = {Handbook of Agricultural Economics},
	publisher = {Elsevier},
	title     = {Chapter 76 - The empirical analysis of climate change impacts and adaptation in agriculture},
	year      = {2021},
	editor    = {Christopher B. Barrett and David R. Just},
	pages     = {3981-4073},
	series    = {Handbook of Agricultural Economics},
	volume    = {5},
	doi       = {https://doi.org/10.1016/bs.hesagr.2021.10.002},
	issn      = {1574-0072},
	url       = {https://www.sciencedirect.com/science/article/pii/S1574007221000025},
}

@Article{Fan2023,
  author   = {Fan, Yichun and Wang, Jianghao and Obradovich, Nick and Zheng, Siqi},
  journal  = {Scientific Reports},
  title    = {Intraday adaptation to extreme temperatures in outdoor activity},
  year     = {2023},
  issn     = {2045-2322},
  number   = {1},
  pages    = {473},
  volume   = {13},
  doi      = {10.1038/s41598-022-26928-y},
  refid    = {Fan2023},
  url      = {https://doi.org/10.1038/s41598-022-26928-y},
}

@Article{Obradovich2017,
  author   = {Obradovich, Nick and Fowler, James H.},
  journal  = {Nature Human Behaviour},
  title    = {Climate change may alter human physical activity patterns},
  year     = {2017},
  issn     = {2397-3374},
  number   = {5},
  pages    = {0097},
  volume   = {1},
  doi      = {10.1038/s41562-017-0097},
  refid    = {Obradovich2017},
  url      = {https://doi.org/10.1038/s41562-017-0097},
}

@Article{Dundas2020,
  author   = {Dundas, Steven J. and von Haefen, Roger H.},
  journal  = {Journal of the Association of Environmental and Resource Economists},
  title    = {The Effects of Weather on Recreational Fishing Demand and Adaptation: Implications for a Changing Climate},
  year     = {2020},
  number   = {2},
  pages    = {209-242},
  volume   = {7},
  doi      = {10.1086/706343},
  eprint   = {https://doi.org/10.1086/706343},
  url      = {https://doi.org/10.1086/706343},
}

@Article{BerrangFord2021,
  author   = {Berrang-Ford, Lea and Siders, A. R. and Lesnikowski, Alexandra and Fischer, Alexandra Paige and Callaghan, Max W. and Haddaway, Neal R. and Mach, Katharine J. and Araos, Malcolm and Shah, Mohammad Aminur Rahman and Wannewitz, Mia and Doshi, Deepal and Leiter, Timo and Matavel, Custodio and Musah-Surugu, Justice Issah and Wong-Parodi, Gabrielle and Antwi-Agyei, Philip and Ajibade, Idowu and Chauhan, Neha and Kakenmaster, William and Grady, Caitlin and Chalastani, Vasiliki I. and Jagannathan, Kripa and Galappaththi, Eranga K. and Sitati, Asha and Scarpa, Giulia and Totin, Edmond and Davis, Katy and Hamilton, Nikita Charles and Kirchhoff, Christine J. and Kumar, Praveen and Pentz, Brian and Simpson, Nicholas P. and Theokritoff, Emily and Deryng, Delphine and Reckien, Diana and Zavaleta-Cortijo, Carol and Ulibarri, Nicola and Segnon, Alcade C. and Khavhagali, Vhalinavho and Shang, Yuanyuan and Zvobgo, Luckson and Zommers, Zinta and Xu, Jiren and Williams, Portia Adade and Canosa, Ivan Villaverde and van Maanen, Nicole and van Bavel, Bianca and van Aalst, Maarten and Turek-Hankins, Lynée L. and Trivedi, Hasti and Trisos, Christopher H. and Thomas, Adelle and Thakur, Shinny and Templeman, Sienna and Stringer, Lindsay C. and Sotnik, Garry and Sjostrom, Kathryn Dana and Singh, Chandni and Siña, Mariella Z. and Shukla, Roopam and Sardans, Jordi and Salubi, Eunice A. and Safaee Chalkasra, Lolita Shaila and Ruiz-Díaz, Raquel and Richards, Carys and Pokharel, Pratik and Petzold, Jan and Penuelas, Josep and Pelaez Avila, Julia and Murillo, Julia B. Pazmino and Ouni, Souha and Niemann, Jennifer and Nielsen, Miriam and New, Mark and Nayna Schwerdtle, Patricia and Nagle Alverio, Gabriela and Mullin, Cristina A. and Mullenite, Joshua and Mosurska, Anuszka and Morecroft, Mike D. and Minx, Jan C. and Maskell, Gina and Nunbogu, Abraham Marshall and Magnan, Alexandre K. and Lwasa, Shuaib and Lukas-Sithole, Megan and Lissner, Tabea and Lilford, Oliver and Koller, Steven F. and Jurjonas, Matthew and Joe, Elphin Tom and Huynh, Lam T. M. and Hill, Avery and Hernandez, Rebecca R. and Hegde, Greeshma and Hawxwell, Tom and Harper, Sherilee and Harden, Alexandra and Haasnoot, Marjolijn and Gilmore, Elisabeth A. and Gichuki, Leah and Gatt, Alyssa and Garschagen, Matthias and Ford, James D. and Forbes, Andrew and Farrell, Aidan D. and Enquist, Carolyn A. F. and Elliott, Susan and Duncan, Emily and Coughlan de Perez, Erin and Coggins, Shaugn and Chen, Tara and Campbell, Donovan and Browne, Katherine E. and Bowen, Kathryn J. and Biesbroek, Robbert and Bhatt, Indra D. and Bezner Kerr, Rachel and Barr, Stephanie L. and Baker, Emily and Austin, Stephanie E. and Arotoma-Rojas, Ingrid and Anderson, Christa and Ajaz, Warda and Agrawal, Tanvi and Abu, Thelma Zulfawu},
  journal  = {Nature Climate Change},
  title    = {A systematic global stocktake of evidence on human adaptation to climate change},
  year     = {2021},
  issn     = {1758-6798},
  number   = {11},
  pages    = {989--1000},
  volume   = {11},
  doi      = {10.1038/s41558-021-01170-y},
  refid    = {Berrang-Ford2021},
  url      = {https://doi.org/10.1038/s41558-021-01170-y},
}

@Article{Carman2020,
  author   = {Jennifer P. Carman and Michaela T. Zint},
  journal  = {Global Environmental Change},
  title    = {Defining and classifying personal and household climate change adaptation behaviors},
  year     = {2020},
  issn     = {0959-3780},
  pages    = {102062},
  volume   = {61},
  doi      = {https://doi.org/10.1016/j.gloenvcha.2020.102062},
  url      = {https://www.sciencedirect.com/science/article/pii/S0959378019306673},
}
